\pdfoutput=1
\documentclass{aa}  
\usepackage{dblfloatfix}
\usepackage{graphicx}
\usepackage{natbib}
\usepackage{capt-of}
\usepackage{textcomp}
\usepackage{epstopdf}
\usepackage{rotating}
\usepackage{siunitx}
\usepackage{multirow}
\usepackage{hhline}
\usepackage{float}
\floatstyle{plaintop}
\restylefloat{table}
\usepackage{threeparttable}
\usepackage{longtable}
\usepackage{import}
\usepackage{array}
\usepackage{amsmath}
\usepackage{mwe}
\usepackage{subfig}
\usepackage[version=4]{mhchem}
\usepackage{xcolor}

\usepackage[colorlinks=true,allcolors=blue]{hyperref}

\begin{document} 

\renewcommand{\thefootnote}{\alph{footnote}}
\renewcommand{\thefootnote}{\fnsymbol{footnote}}

   \title{The hunt for formamide in interstellar ices}
   \subtitle{A toolkit of laboratory infrared spectra in astronomically relevant ice mixtures and comparisons to ISO, Spitzer, and JWST observations}
\titlerunning{}

   \author{K. Slavicinska\inst{1,2}
   \and
    M. G. Rachid\inst{1}
    \and
    W. R. M. Rocha\inst{1,2}
    \and
    K. -J. Chuang\inst{1}
    \and
    E. F. van Dishoeck\inst{2,3}
    \and
    H. Linnartz\inst{1}}

   \institute{Laboratory for Astrophysics, Leiden Observatory, Leiden University, P.O. Box 9513, 2300 RA Leiden, The Netherlands.\\
   \email{slavicinska@strw.leidenuniv.nl}
   \and
   Leiden Observatory, Leiden University, P.O. Box 9513, 2300 RA Leiden, The Netherlands.
   \and
   Max Planck Institut f\"ur Extraterrestrische Physik (MPE), Giessenbachstrasse 1, 85748 Garching, Germany}

   \date{Received 24 May 2023 / Accepted 30 June 2023}



  \abstract
   {Although solid-state pathways are expected to dominate the formation mechanisms of many complex organic molecules (COMs), very few COMs have been securely identified in interstellar ices, in stark contrast with the many COM detections in the gas phase. The launch of the James Webb Space Telescope (JWST) and its increase in sensitivity and spectral resolution opens the possibility of identifying more COMs in ices, but additional laboratory data are necessary. Formamide (NH$_{2}$CHO) is one such COM that is of great interstellar and prebiotic relevance where more laboratory data are needed in the hunt for its presence in interstellar ices.}
   {This work aims to characterize the mid-IR spectra of formamide in its pure form as well as in mixtures of the most abundant interstellar ices via laboratory simulation of such ices, as well as to demonstrate how these laboratory spectra can be used to search for formamide in ice observations.}
   {Mid-IR spectra (4000 - 500 cm$^{-1}$/2.5 - 20 $\mu$m) of formamide, both in its pure form as well as in binary and tertiary mixtures with H$_2$O, CO$_2$, CO, NH$_3$, CH$_3$OH, H$_2$O:CO$_2$, H$_2$O:NH$_3$, CO:NH$_3$, and CO:CH$_3$OH, were collected at temperatures ranging from 15 - 212 K.}
   {Apparent band strengths and positions of eight IR bands of pure amorphous and crystalline formamide at various temperatures are provided. Three of these bands are identified as potential formamide tracers in observational ice spectra: the overlapping C=O stretch and NH$_2$ scissor bands at 1700.3 and 1630.4 cm$^{-1}$ (5.881 and 6.133 $\mu$m), the CH bend at 1388.1 cm$^{-1}$ (7.204 $\mu$m), and the CN stretch at 1328.1 cm$^{-1}$ (7.529 $\mu$m). The relative apparent band strengths, positions, and full width half maxima (FWHM) of these features in mixtures at various temperatures were also determined. All of the laboratory spectra are available to the community on the Leiden Ice Database for Astrochemistry (LIDA) for use in the interpretation of both observations (e.g., from JWST) and laboratory spectroscopic data. Finally, the laboratory spectra are compared to observational spectra of a variety of low- and high-mass young stellar objects as well as prestellar cores observed with the Infrared Space Observatory, the Spitzer Space Telescope, and JWST. A comparison between the formamide CH bend in laboratory data and the 7.24 $\mu$m band in the observations tentatively indicates that, if formamide ice is contributing significantly to the observed absorption, it is more likely in a polar matrix. Upper limits ranging from 0.35-5.1\% with respect to H$_{2}$O were calculated via scaling the formamide:H$_{2}$O laboratory spectrum to the observations. These upper limits are in agreement with gas-phase formamide abundances and take into account the effect of a H$_{2}$O matrix on formamide's band strengths.}
   {}


   \maketitle
%
\section{Introduction}

Of the $>$280 molecules that have been detected in interstellar environments \citep{endres2016cologne}, formamide (NH$_2$CHO) has become one of the most widely and deeply investigated in observational, modeling, computational, and laboratory studies in the last decade. Containing all four of the most abundant biological elements (C, H, N, and O), formamide is the simplest molecule that contains the biologically essential amide bond and has been suggested as a plausible prebiotic precursor to various nucleobases (e.g., \citealt{saladino2003one,barks2010guanine}), the chemical building blocks of RNA and DNA. It has also been proposed as an alternative prebiotic solvent to promote condensation reactions, which form many vital biological molecules but are highly endergonic in purely aqueous solutions (e.g., phosphorylation), by lowering water activity \citep{gull2017silicate,pasek2019thermodynamics,lago2020prebiotic}.

Given this potential prebiotic relevance, the fact that formamide has been observed in numerous sources in the interstellar medium as well as on extraterrestrial bodies in our own Solar System has exciting implications for astrobiology. First detected in the interstellar medium in the gas phase by \citet{rubin1971microwave} in the Sagittarius B2 high-mass star-forming region, formamide has since been observed in over 30 massive young stellar objects (MYSOs) as well as low-mass YSOs (LYSOs) with hot corinos and protostellar shocks (\citealt{lopez2019interstellar} and references therein). Within our Solar System, gas-phase formamide has been found in the comae of the comets Lemmon, Lovejoy, and Hale-Bopp, with abundances ranging around 0.01-0.02\% with respect to H$_2$O \citep{bockelee2000new,biver2014complex}. It was also detected in situ by the Rosetta mission on comet 67P Churyumov-Gerasimenko, both on the surface by the Cometary Sampling and Composition experiment (COSAC) instrument on the Philae lander \citep{goesmann2015organic} and in the coma by the Double Focusing Mass Spectrometer (DFMS) on the Rosetta Orbiter Spectrometer for Ion and Neutral Analysis (ROSINA) instrument \citep{altwegg2017organics,rubin2019elemental}, where the formamide abundance was found to be $\sim$0.004\% with respect to H$_2$O.

Notably, all of the interstellar sources in which gas-phase formamide has been securely detected have hot cores and corinos or shocked regions, where temperatures are high enough for formamide to thermally desorb from icy grains into the gas phase \citep{lopez2019interstellar}. Additionally, in many of these sources, the formamide abundance correlates almost linearly with the abundance of isocyanic acid (HNCO) \citep{bisschop2007testing,mendoza2014molecules,lopez2015shedding}, and, in the case of the low-mass source IRAS 16293-2422, the two species are spatially correlated and have very similar deuteration ratios \citep{coutens2016alma}.

These aspects of formamide observations could be considered evidence that formamide is formed in the solid state (i.e., via ice chemistry), possibly in a pathway chemically related to HNCO, and it is detected in the gas phase following desorption from icy grains. The ice formation and grain sublimation scenario is further supported by recent observational work investigating excitation temperatures of N-bearing complex organic molecules (COMs) in 37 MYSOs from the ALMA Evolutionary study of High Mass Protocluster Formation in the Galaxy (ALMAGAL) survey, where formamide had the highest excitation temperatures of all the studied N-bearing COMs ($\gtrsim$250 K) \citep{nazari2022n}. These temperatures are consistent with thermal desorption experiments, in which formamide ice sublimes at high temperatures (typically >210 K) even when it is mixed with or deposited on top of more volatile species such as H$_2$O and CO, and at even higher temperatures (>250 K) when the experiments are performed on certain dust grain analog substrates \citep{dawley2014thermal,urso2017infrared,chaabouni2018thermal,corazzi2020photoprocessing}.

\begin{figure}
\centering
\includegraphics[width=0.55\linewidth]{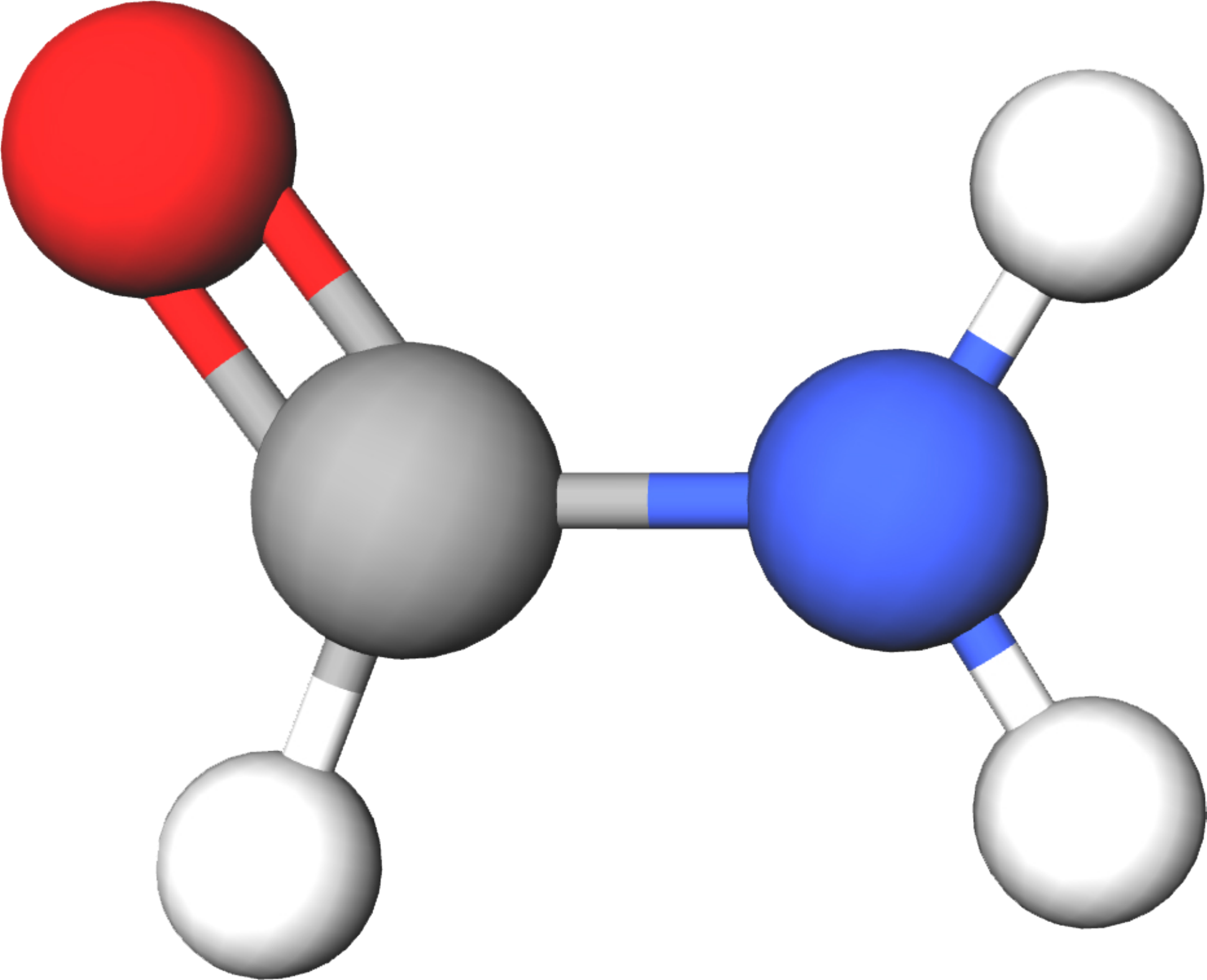}
\caption{Molecular structure of NH$_2$CHO (hydrogen, white; carbon, gray; nitrogen, blue; and oxygen, red).}
\label{fig:nh2cho}
\end{figure}

Experimentally, solid-state formamide has been identified as a product of processing via a variety of energetic sources (e.g., electron, UV, X-ray, and ion irradiation) of a myriad of simple ice mixtures, including (but not limited to) CO:NH$_3$ \citep{demyk1998laboratory,hudson2000new,jones2011mechanistical,bredehoft2017electron,martin2020formation} and H$_2$O:CO:NH$_3$ \citep{hudson2000new,ciaravella2019synthesis,chuang2022formation}, NO:H$_2$CO:H and NO:CH$_3$OH:H \citep{fedoseev2016simultaneous}, H$_2$O:HCN \citep{gerakines2004ultraviolet}, H$_2$O:CH$_4$:NH$_3$ and H$_2$O:CH$_4$:N$_2$ \citep{kavnuchova2016synthesis}, and HNCO \citep{raunier2004tentative} and CH$_4$:HNCO \citep{ligterink2018formation}. Evidently, energetic processing of almost any ice mixture that contains H, N, C, and O is very likely to produce formamide. Such processing experiments mimic the radiation environments experienced by ices in protostellar envelopes and protoplanetary disks. Furthermore, recent experiments by \citet{dulieu2019efficient} demonstrate that hydrogenation of NO:H$_2$CO can also produce formamide, providing a plausible nonenergetic formation pathway that is relevant to cold, dark clouds.

While a plethora of observational, experimental, and theoretical works (see Section~\ref{debate}) have significantly progressed our understanding of formamide's interstellar presence and its plausible chemical history, whether its formation occurs in the solid state, gas phase, or both remains unclear. A secure detection of formamide in ices would be immensely valuable to resolve this debate regarding its formation mechanism. Such a detection, if well resolved, could provide parameters such as formamide's solid-state abundance and its physico-chemical environment, which are essential to elucidating its formation pathway.

Previously, formamide has been tentatively detected in the solid state in the Infrared Space Observatory Short Wavelength Spectrometer (ISO-SWS) spectra of the MYSOs W33A and NGC 7538 IRS 9. In the case of W33A, an upper limit of 2.1\% with respect to H$_2$O via the CH bend at 7.22 $\mu$m/1385 cm$^{-1}$ was derived, but the authors noted that the peak position in the observation (7.24 $\mu$m) was red-shifted relative to the formamide peak in their laboratory spectra \citep{schutte1999weak}. For NGC 7538 IRS 9, no upper limit of formamide was provided -- a laboratory spectrum of irradiated HNCO that showed IR evidence of formamide formation was qualitatively evaluated as a spectral fit to the observed 6 $\mu$m/1700 cm$^{-1}$ band \citep{raunier2004tentative}. In both of these cases, the bands attributed to formamide were overlaid on top of or blended with other strong ice features.

Typically, reference laboratory IR spectra are used to assign and fit astronomically observed IR features to specific species, and band strengths acquired via systematic laboratory experiments are used to quantify the column densities of these species. For COMs such as formamide that are expected to be present in the ice in very low concentrations ($\lesssim$5\%), it is important to obtain these spectra and band strengths not only for pure ices, but also in chemical conditions that are more realistic for interstellar ices. Namely, the molecule of interest should be diluted in the more abundant simple ice species (e.g., H$_{2}$O, CO, and CO$_{2}$), as interactions with other species present in the ice matrix can significantly alter the positions, profiles, and apparent band strengths of a molecule's vibrational features. Morphological changes in the ice caused by thermal processing, such as transitions from amorphous to crystalline ice or matrix segregation, can also dramatically change an ice's spectral features, so spectra should be collected at a variety of temperatures as well. Considering such factors is not only important to accurately assign and quantify the molecule of interest, but it can also provide valuable information about the molecule's physico-chemical environment and history.

In previous IR characterization work, \citet{brucato2006infrared} derived the refractive index, density, and several band strengths of pure formamide, but integration ranges and errors were not provided for these band strengths, and no spectra of heated formamide or formamide in mixtures were collected. In order to tentatively assign the 7.24 $\mu$m band in W33A's spectrum to formamide, \citet{schutte1999weak} collected spectra of formamide at 10 K in H$_{2}$O and H$_{2}$O:CH$_{3}$OH matrices, but only one band was characterized from these spectra, and it is unclear for what phase of formamide the band strength used in the upper limit calculation was derived. \citet{urso2017infrared} collected IR spectra of formamide in pure, H$_{2}$O-dominated, and CO-dominated ice matrices, but the band strengths, peak positions, and full width half maxima (FWHMs) of the formamide features in these mixtures are not given. \citet{sivaraman2013infrared} presented the peak positions of the bands of pure formamide in the 30 - 210 K temperature range, but no spectra of formamide in mixtures were collected.

Thus, in an effort to enable more secure assignments and accurate abundance and/or upper limit determinations of formamide in observed ice spectra, this work provides a comprehensive set of laboratory transmission IR spectra of pure formamide as well as formamide diluted in nine different astrophysically relevant ice mixtures of varying polarities. These spectra are provided at temperatures ranging from 15 - 212 K. Apparent band strengths were derived for eight integrated regions from the pure formamide spectra, and from these, three bands are evaluated as the most promising for future identification of formamide in observations. These bands are also fully characterized (i.e., peak positions, FWHMs, and relative band strengths are provided). Examples of how these spectra and values can be used in future analyses of ice observations are described, and new upper limits of formamide in a variety of objects (prestellar cores, low-mass protostars, and high-mass protostars) were calculated. Finally, all spectra are made publicly available on the Leiden Ice Database\footnote{https://www.icedb.strw.leidenuniv.nl} \citep{rocha2022lida} for the community to use in fitting to their ice observations. This work is particularly timely given the recent launch of the James Webb Space Telescope (JWST), which may enable the detection of new COMs in interstellar ices due to its unprecedented sensitivity and spectral resolution.

\section{Formamide formation mechanism debate}
\label{debate}

A variety of pathways have been suggested to explain the observed solid-state formamide formation in laboratory ice experiments. One initially proposed mechanism was the hydrogenation of HNCO, an attractive premise given that it provided a direct chemical link between HNCO and formamide to explain their correlation in gas-phase observations:\\

\ce{HNCO + 2^.H \rightarrow NH2CHO.}\\

This pathway was first suggested by \citet{charnley1997astronomical} and was stated as a possible formation mechanism of formamide when it was observed in VUV irradiation experiments of pure HNCO \citep{raunier2004tentative}. However, hydrogenation experiments by \citet{noble2015hydrogenation} via H bombardment of HNCO <20 K did not produce detectable amounts of formamide, although the authors suggested that the reaction may be prevented in their experiments by the formation of very stable HNCO dimers or polymers, and that it could possibly proceed if HNCO is diluted in the matrix of an ice like H$_2$O. Indeed, subsequent experiments by \citet{haupa2019hydrogen} showed that, in a 3.3 K para-H$_2$ matrix, formamide can form from HNCO via a hydrogen addition-abstraction cycling mechanism, but in this reaction scheme, HNCO is still the favored product.

Another proposed formation pathway is the following radical-radical recombination:\\

\ce{^.NH2 + ^.CHO \rightarrow NH2CHO.}\\

\noindent This mechanism is technically barrierless and can proceed at low temperatures ($\sim$10 K) but produces higher yields at higher temperatures ($\sim$20-40 K) due to increased mobility allowing the radicals to orient in the proper reaction geometry \citep{rimola2018can,martin2020formation}. In the laboratory, this mechanism requires some form of energetic processing to generate the NH$_{2}$ and CHO radicals, and its viability is supported by the presence of the CHO radical in the experimental spectra \citep{jones2011mechanistical,fedoseev2016simultaneous,ciaravella2019synthesis,martin2020formation,chuang2022formation}.

Various mechanisms have also been suggested where formamide is produced from the NH$_{2}$CO radical, which could form by the radical-molecule association of NH$_{2}$ and CO or CN and H$_{2}$O \citep{hudson2000new,bredehoft2017electron,rimola2018can}:\\

\ce{^.NH2CO + ^.H \rightarrow NH2CHO}\\
\indent \ce{^.NH2CO + H2O \rightarrow NH2CHO + ^.OH}\\
\indent \ce{2^.NH2CO \rightarrow NH2CHO + HNCO.}\\

\noindent However, the formation of the NH$_{2}$CO radical via a pathway that does not involve hydrogen abstraction from already existing formamide, as seen in \citet{haupa2019hydrogen}, has yet to be experimentally confirmed.

While these latter mechanisms do not provide an immediately obvious direct solid-state link between HNCO and NH$_{2}$CHO, some experimental studies have suggested alternative links consistent with these mechanisms. For example, once formed, formamide can decompose into HNCO via dehydrogenation and photolysis by H$_{2}$ loss \citep{brucato2006infrared,haupa2019hydrogen,chuang2022formation}, so HNCO may be a product of NH$_2$CHO rather than the other way around. \citet{fedoseev2016simultaneous} proposed that the NH$_{2}$ radical can produce either HNCO or NH$_2$CHO depending on the degree of hydrogenation of the C- and O-containing molecule with which it reacts: the reaction of NH$_{2}$ with CO leads to HNCO, while NH$_{2}$ with HCO or H$_{2}$CO leads to formamide.

Thus, while formamide may not be a direct product of HNCO, the two species may be linked in a solid-state chemical network by common precursors. Astrochemical models using the rate constants from \citet{fedoseev2016simultaneous} further corroborate that, indeed, a direct chemical link between HNCO and NH$_2$CHO is not necessary to reproduce the observed linear correlation between them in models of various interstellar environments and suggest instead that their correlation could be explained by their similar responses to physical (i.e., thermal) environments \citep{quenard2018chemical}.

In addition to these solid-state mechanisms, the plausibility of the following gas-phase formation route has been extensively debated in computational and modeling works since its proposal in \citet{garrod2008complex}:\\

\ce{^.NH2 + H2CO \rightarrow NH2CHO + ^.H.}\\

According to its first published electronic structure and kinetic calculations, this reaction is essentially barrierless at low temperatures and thus should proceed readily in interstellar environments \citep{barone2015gas,vazart2016state}. Furthermore, chemical models of the protostar IRAS 16293-2422 and the molecular shocks L1157-B1 and B2 utilizing the calculated rate coefficients of this reaction produce formamide abundances that are consistent with observed values \citep{barone2015gas,codella2017seeds}, and follow-up studies calculating rate coefficients of deuterated formamide formation via the same reaction show that formamide's observed deuteration ratio does not necessarily exclude the possibility of gas-phase formation \citep{skouteris2017new}.

However, the accuracy of these calculated rate coefficients has been called into question given that they neglect the zero point energy (ZPE) of one of the transition states. When the ZPE of the transition state is included, the reaction barrier becomes large enough that the reaction rate is negligible at low temperatures \citep{song2016formation}, although some argue that inclusion of the ZPE is not warranted for this transition state and results in overestimation of the reaction barrier \citep{skouteris2017new}. Recent gas-phase experiments attempting to perform this route did not confirm any formamide formation, and their detection upper limits are consistent with the reaction barrier that includes the transition state ZPE \citep{douglas2022gas}.

\section{Methodology}
\label{section:methodology}

All of the measurements were collected in the Laboratory for Astrophysics at Leiden Observatory on the IRASIS (InfraRed Absorption Setup for Ice Spectroscopy) chamber. The setup was described in detail in \citet{rachid2021infrared} and \citet{rachid2022infrared}, and it has since undergone several upgrades, including a decrease of its base pressure to <1.0$\times$10$^{-9}$ mbar by the addition of new pumps, an exchange of the laser used for interference measurements to one with a wavelength of 543 nm (as the formamide ice refractive index was measured by \citealt{brucato2006infrared} at this wavelength), and the implementation of an independent tri-dosing leak valve system that can be calibrated with a quadrupole mass spectrometer (QMS) following the procedure described in Appendix~\ref{supp_info_qms}.

The optical layout of the chamber remains the same as that shown in Figure 1 in \citet{rachid2021infrared}: a Ge substrate sits at the center of the chamber and is cooled by a closed-cycle He cryostat to 15 K. Ices are grown on the substrate via background deposition of gases and vapors dosed into the chamber through leak valves. Infrared transmission spectra are collected through two ZnSe viewports that are parallel to the Ge substrate and normal to the IR light beam. During deposition, laser interference patterns used to determine ice thickness are measured on both sides of the Ge substrate (which is opaque and reflective in the visible light range) via photodiode detectors placed outside of viewports positioned 45$^{\circ}$ from the substrate normal. The patterns obtained from each side of the substrate during deposition show equal deposition rates on both sides. After deposition, the substrate can be heated to obtain IR spectra at different temperatures. In this work, 256 spectral scans with a 0.5 cm$^{-1}$ resolution were collected and averaged while the substrate was heated at a rate of 25 K hr$^{-1}$, resulting in a temperature uncertainty of $\pm$1.5 K in each heated spectrum. Spectra were collected during heating until reaching the temperature at which the major matrix component desorbed. Before their analysis, all spectra were baseline-corrected using a cubic spline function.

The liquids and gases used in this work were formamide (Sigma Aldrich, $\ge$99.5\%), water (Milli-Q, Type I), carbon dioxide (Linde, $\ge$99.995\%), carbon monoxide (Linde, $\ge$99.997\%), ammonia (PraxAir, $\ge$99.96\%), and methanol (Sigma Aldrich, $\ge$99.9\%). The mixing ratios calculated for all of the spectra via the method outlined in Appendix~\ref{supp_info_qms} are presented in Table~\ref{ratios}. Uncertainties in the column densities used to calculate these ratios are estimated to be $\sim$21\% for the formamide column densities and $\sim$27\% for the matrix species column densities (see Appendix~\ref{supp_info_qms}). Prior to deposition, the liquid formamide sample was heated to 60$^{\circ}$C and pumped on directly with a turbomolecular pump in order to remove contaminants (primarily water).

\begin{table}[h]
\caption{Experimentally determined ratios of all mixtures presented in this work.}
\begin{center}
\begin{tabular}{c c}
\hline
        Mixture & Ratio \\
        \hline
        Pure NH$_{2}$CHO & -- \\
        NH$_{2}$CHO:H$_{2}$O & 7:100 \\
        NH$_{2}$CHO:CO$_{2}$ & 7:100 \\
        NH$_{2}$CHO:CO & 4:100 \\
        NH$_{2}$CHO:CH$_{3}$OH & 8:100 \\
        NH$_{2}$CHO:NH$_{3}$ & 7:100 \\
        NH$_{2}$CHO:H$_{2}$O:CO$_{2}$ & 8:100:37 \\
        NH$_{2}$CHO:H$_{2}$O:NH$_{3}$ & 8:100:31 \\
        NH$_{2}$CHO:CO:CH$_{3}$OH & 4:100:12 \\
        NH$_{2}$CHO:CO:NH$_{3}$ & 5:100:16 \\
    \hline
     \label{ratios}
\end{tabular}
\end{center}
\end{table}

\begin{table*}[h]
\caption{Apparent IR band strengths (and their peak positions, assignments, and corresponding integrated regions) of the IR bands of amorphous NH$_2$CHO ice at 15 K derived in this work.}
\begin{center}
\begin{tabular}{c  c  c  c  c}
\hline
         Peak positions & Integrated region & Assignment$^{*}$  & A' & A'$^{**}$ \\
         (cm$^{-1}$ | $\mu$m) & (cm$^{-1}$) &  & \multicolumn{2}{c}{(10$^{-17}$ cm molec$^{-1}$)} \\
        \hline 
        \vspace{1mm}
         \begin{tabular}{c}3371.2 | 2.966 \\3176.4 | 3.148 \end{tabular} & 3719 - 2980  & \begin{tabular}{c}NH$_2$ antisymm. stretch ($\nu_{as}$ NH$_{2}$) \\NH$_2$ symm. stretch ($\nu_{s}$ NH$_{2}$) \end{tabular} & 15.6 $\pm$ 2 & 13.49\\
         \vspace{1mm}
         2881.9 | 3.470 & 2980 - 2842 & CH stretch ($\nu$ CH) & 0.56 $\pm$ 0.08 & 0.47 \\
         \vspace{1mm}
         2797.7 | 3.574 & 2842 - 2600 & CH bend overtone (2$\delta$ CH) & 0.23 $\pm$ 0.03 & -- \\
         \vspace{1mm}
         \begin{tabular}{c}1700.3 | 5.881 \\1630.4 | 6.133 \end{tabular} & 2050 - 1532 & \begin{tabular}{c}C=O stretch ($\nu$ CO) \\NH$_2$ scissor ($\delta$ NH$_{2}$) \end{tabular}  & 6.4 $\pm$ 0.9$^{***}$ & 6.54 \\
         \vspace{1mm}
         1388.1 | 7.204 & 1532 - 1364 & CH bend ($\delta$ CH) & 0.9 $\pm$ 0.1$^{***}$ & 0.68 \\
        \vspace{1mm}
         1328.1 | 7.529 & 1364 - 1258 & CN stretch ($\nu$ CN) & 0.8 $\pm$ 0.1$^{***}$ & 0.85 \\
         \vspace{1mm}
         \begin{tabular}{c}1108.1 | 9.024 \\1056.1 | 9.469 \end{tabular} & 1174 - 1030 & \begin{tabular}{c}NH$_2$ rock ($\rho$ NH$_{2}$) \\CH out-of-plane def. ($\gamma$ CH) \end{tabular} & 0.23 $\pm$ 0.03 & -- \\
         \begin{tabular}{c}689.2 | 14.510 \\634.0 | 15.773 \end{tabular} & 1030 - 575 & \begin{tabular}{c}NH$_2$ wag ($\omega$ NH$_{2}$)\\ NH$_2$ twist (2$\tau$ NH$_{2}$) \end{tabular} & 2.3 $\pm$ 0.4 & -- \\

        \hline
\end{tabular}

\begin{tablenotes}
    \item[\emph{}]{$^*$Band assignments made according to the works of \citealt{rasanen1983matrix,kwiatkowski1993molecular,torrie1994raman,lundell1998matrix}. Some band strengths contain multiple absorption modes in their corresponding integrated regions.}
    \item[\emph{}]{$^{**}$Values from \citealt{brucato2006infrared}.}
    \item[\emph{}]{$^{***}$Bands that are also characterized in mixtures in this work.}

\end{tablenotes}
\end{center}

\label{band-strength}
\end{table*}

\begin{figure*}[ht]
\centering
\includegraphics[scale=0.7]{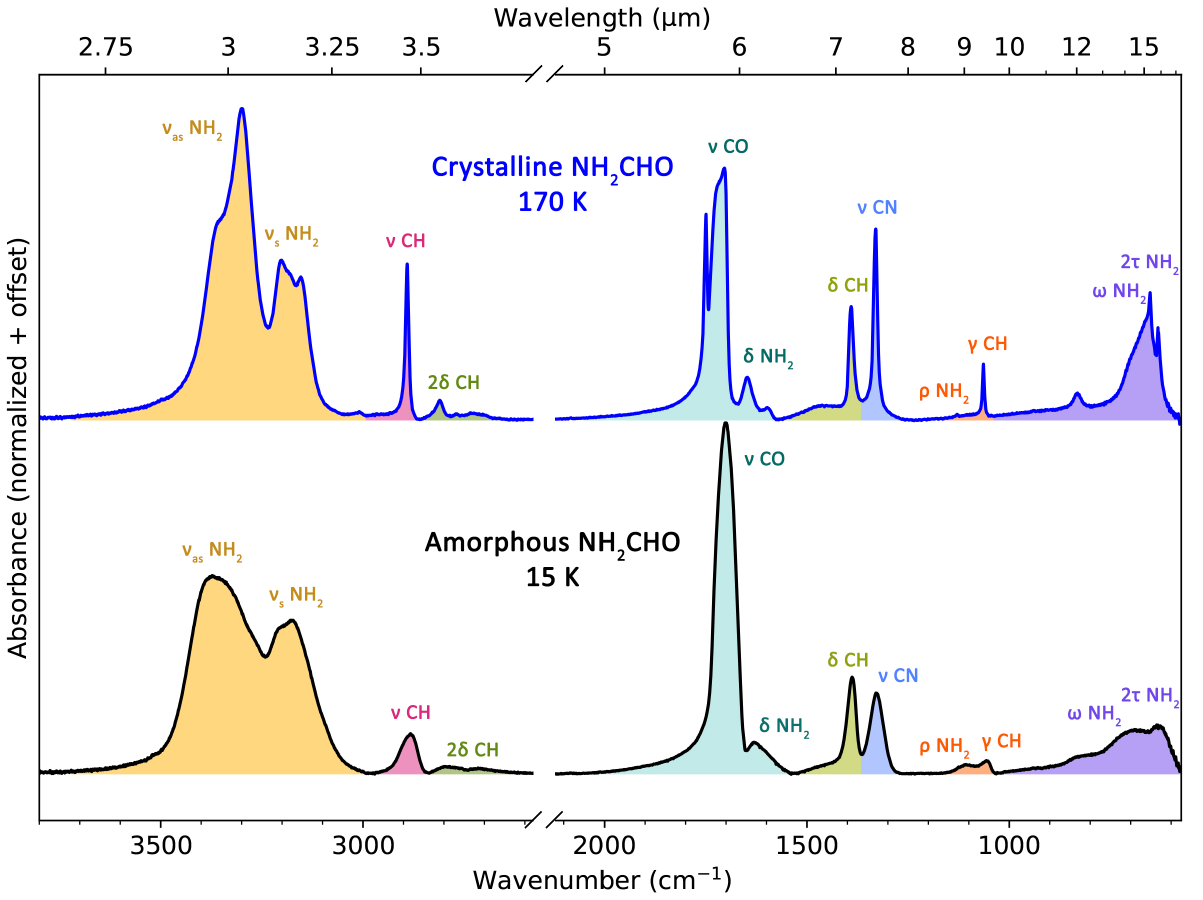}
\caption{Infrared spectra in the range 3800 - 575 cm$^{-1}$ of both amorphous and crystalline formamide (NH$_{2}$CHO). The integrated regions used to calculate the band strengths in Table~\ref{band-strength} are indicated via the shaded areas under the peaks.}
\label{fig:matrix_spectra}
\end{figure*}

The apparent band strengths of pure formamide are determined via depositing formamide onto the substrate held at 15 K while simultaneously collecting the transmission IR spectra and the laser interference pattern. The thickness \textit{d} of the ice can be derived from the laser interference pattern via the following equation:

\begin{equation}
    \indent d = \frac{m\lambda}{2\sqrt{n^{2} - sin^{2}\theta}},
\end{equation}

\noindent where \textit{m} is an integer number of constructive fringes, \textit{$\lambda$} is the laser wavelength, \textit{n} is the ice refractive index (1.361 for formamide at 543 nm, from \citealt{brucato2006infrared}), and \textit{$\theta$} is the angle of incidence.

Enough formamide is deposited so that four constructive fringes are acquired, the thickness of the ice at each fringe peak is calculated, and the integrated absorbances of eight spectral regions (see Table~\ref{band-strength}) are calculated from the spectra collected at the time that a fringe peak was reached. Then, the integrated absorbance for each spectral region is plotted as a function of ice thickness, and the slope of this line, $\Delta\int abs(\nu) \ d\nu/\Delta d$, is obtained via a least-squares fit. From this value, the apparent band strengths \textit{A'} can be approximated with an equation based on the Beer-Lambert Law (e.g., \citealt{hudson2014infrared,gerakines2023carbon}):

\begin{equation}
    \indent A' = \frac{2.303 \; M}{\rho \; N_{A}} \times \frac{\Delta\int abs(\nu) \ d\nu}{\Delta d},
\label{eq:aval}
\end{equation}

\noindent where \textit{M} is the molar mass of formamide (45.041 g mol$^{-1}$), \textit{$\rho$} is the density of formamide ice (0.937 g cm$^{-3}$, from \citealt{brucato2006infrared}), and \textit{N$_{A}$} is Avogadro's number. Using change in integrated absorbance over change in thickness in this equation rather than the absolute values of both variables ensures that there is no contribution of any residue from previous experiments on the substrate to the calculated ice thickness. It also does not require a constant ice growth rate.

The apparent band strengths reported in Table~\ref{band-strength} are the averages of three repeated measurements following this method. The experimental uncertainties derived from the standard deviation of these three measurements range from 3-8\% for the eight band strengths. However, simply using the standard deviations from the repeated measurements as the band strengths uncertainties neglects potential systemic sources of error such as uncertainties in the laser alignment geometry and the data analysis procedure. Thus, the uncertainties provided in Table~\ref{band-strength} are calculated via error propagation of all of the experimental terms in Equation~\ref{eq:aval}, using the same estimated uncertainties as \citet{rachid2022infrared} for the ice thickness (4\%) and integrated absorbance (10\%) as well as the ice density (10\%). This calculation yields an uncertainty of 15\% for the reported band strength values.


From the pure formamide apparent band strengths, the apparent band strengths of formamide in the investigated mixtures, \textit{A'$_{i}$}, are calculated using the formamide column densities \textit{N$_{mix}$} (obtained from the methods described in Appendix~\ref{supp_info_qms}) via the following equation:

\begin{equation}
    \indent A'_{i} = 2.303 \times \frac{\int abs(\nu) \ d\nu}{N_{mix}},
\end{equation}

\noindent and the relative apparent band strengths, \textit{$\eta$}, are subsequently found by:

\begin{equation}
    \indent \eta = \frac{A'_{i}}{A'},
\label{eq:relbs}
\end{equation}

Following propagation of error from the pure apparent band strengths, integrated absorbances, and the formamide column densities in the mixtures (see Appendix~\ref{supp_info_qms}), the uncertainties of the relative apparent band strengths presented here are estimated to be $\sim$28\%.

\section{Results}

The spectra of pure amorphous and crystalline formamide are presented in Figure~\ref{fig:matrix_spectra}, and the eight apparent band strengths calculated at 15 K are presented in Table~\ref{band-strength}. Peak positions and vibrational mode assignments are also provided. Some integrated regions contain multiple overlapping peaks; in these cases, the peak positions and assignments were provided for all peaks within the integrated region, but the peaks were not deconvolved to give an individual band strength for each peak. These band strengths have percent differences ranging from 1-35\% compared to those given for the same peak values in \citet{brucato2006infrared}. As integration bounds were not provided by \citet{brucato2006infrared}, any discrepancies in band strengths may be caused by differences in chosen integration regions.

The transition from amorphous to crystalline formamide is observed at 170 K, indicated by its bands becoming sharper and narrower and some peaks splitting. The amorphous nature of almost all of the pure and mixed ices collected at 15 K can be ascertained from their spectra, which have typical amorphous features that show evidence of matrix crystallization during the warm-up phase of the experiments. This excludes the mixtures containing CO, whose phase at 15 K in these experiments may be crystalline given recent investigations of CO ice structure $\ge$10 K \cite{he2021phase,gerakines2023carbon,rachid2023morphological}.

\begin{figure*}[ht]
\includegraphics[width=\hsize]{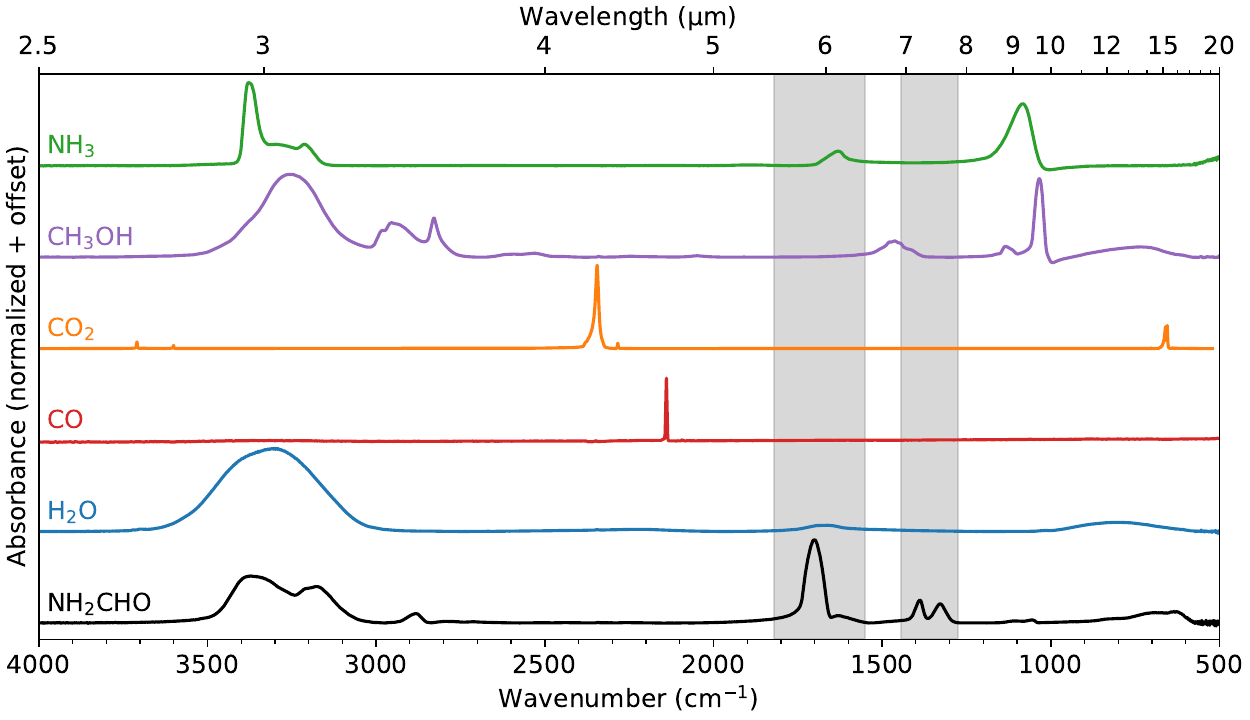}
\caption{Infrared spectra of pure formamide (NH$_{2}$CHO) and five molecules observed in high abundances in interstellar ices that are present in the mixtures analyzed in this work (H$_{2}$O, CO, CO$_{2}$, CH$_{3}$OH, NH$_{3}$), all at 15 K. Peaks selected for further characterization in mixtures are indicated in the shaded areas.}
\label{fig:matrix_spectra}
\end{figure*}

\begin{figure*}
\centering
\includegraphics[width=0.95\textwidth]{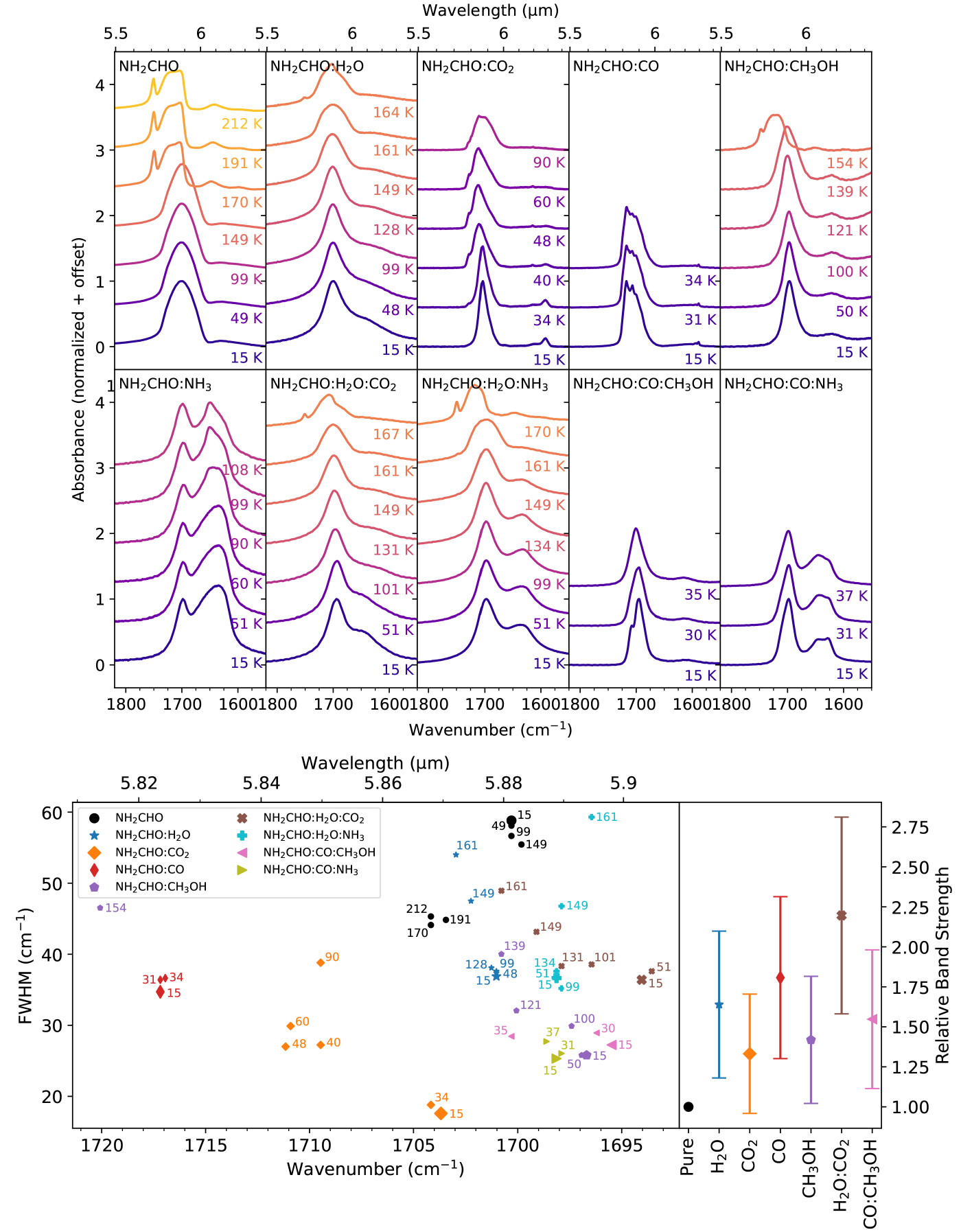}
\caption{Characterization of the C=O stretching and NH$_2$ scissoring bands of NH$_2$CHO. Top: the C=O stretching and NH$_2$ scissoring bands of NH$_2$CHO in pure and mixed forms at selected temperatures. Bottom left: the peak position and FWHM of the formamide C=O stretch in pure and mixed forms at selected temperatures. Bottom right: the band strengths of the formamide C=O stretch and NH$_{2}$ scissor in mixtures relative to the pure band strength (all at 15 K).}
\label{fig:co_stretch}
\end{figure*}

Figure~\ref{fig:matrix_spectra} presents the spectrum of pure formamide ice along with the spectra of the pure matrix components, all at 15 K. The formamide peaks indicated in the shaded areas were selected for full characterization (i.e., their peak positions, FWHMs, and relative band strengths are determined for mixtures): the overlapping C=O stretch and NH$_{2}$ scissor at 1700.3 cm$^{-1}$/5.881 $\mu$m and 1630.4 cm$^{-1}$/6.133 $\mu$m, respectively, and the slightly overlapping CH bend and CN stretch at 1388.2 cm$^{-1}$/7.204 $\mu$m and 1328.1 cm$^{-1}$/7.529 $\mu$m, respectively. These peaks were selected because they are strong, have sharp profiles, and overlap the least with the major peaks of the most common interstellar ices, making them the best candidates for identifying formamide in interstellar ice spectra. There is still some overlap between these formamide peaks and some minor peaks of the matrix components, namely the water OH bend at $\sim$1600 cm$^{-1}$/6.25 $\mu$m, the methanol CH$_{3}$ and OH bends at $\sim$1460 cm$^{-1}$/6.85 $\mu$m, and the ammonia NH scissoring at 1624 cm$^{-1}$/6.16 $\mu$m. However, with sufficiently high formamide concentrations, it may still be possible to identify formamide in these spectral regions, as these matrix bands are relatively weak and broad.

The matrix- and temperature-dependent changes in these selected formamide ice bands are discussed in the following subsections, and their peak positions, FWHMs, and relative band strengths in different mixtures at various temperatures are reported in Appendices~\ref{appendix_a} and~\ref{appendix_b}. The NH$_{2}$ stretching features at 3371.2 cm$^{-1}$/2.966 $\mu$m and 3176.4 cm$^{-1}$/3.148 $\mu$m and the NH$_{2}$ wagging and twisting features at 689.2 cm$^{-1}$/14.510 $\mu$m and 634.0 cm$^{-1}$/15.773 $\mu$m were excluded from further characterization despite their relatively large band strengths due to their direct overlap with the two most intense water features, the OH stretch at $\sim$3279 cm$^{-1}$/3.05 $\mu$m and the H$_{2}$O libration at $\sim$780 cm$^{-1}$/12.8 $\mu$m, respectively \citep{oberg2007effects}. The remaining formamide bands, the CH stretch at 2881.9 cm$^{-1}$/3.470 $\mu$m, the CH bend overtone at 2797.7 cm$^{-1}$/3.574 $\mu$m, and the convolved NH$_{2}$ rock at 1108.1 cm$^{-1}$/9.024 $\mu$m and CH out-of-plane deformation at 1056.1 cm$^{-1}$/9.469 $\mu$m, have low band strengths and directly overlap with various methanol features: the CH$_{3}$ stretches at 2950 cm$^{-1}$/3.389 $\mu$m and 2830 cm$^{-1}$/3.533 $\mu$m, the CH$_{3}$ rock at 1126 cm$^{-1}$/8.881 $\mu$m, and the C-O stretch at 1027 cm$^{-1}$/9.737 $\mu$m \citep{luna2018densities}.

\subsection{C=O stretching and NH$_2$ scissoring features ($\sim$1700 and 1630 cm$^{-1}$)}

\begin{figure*}
\centering
\includegraphics[width=0.95\textwidth]{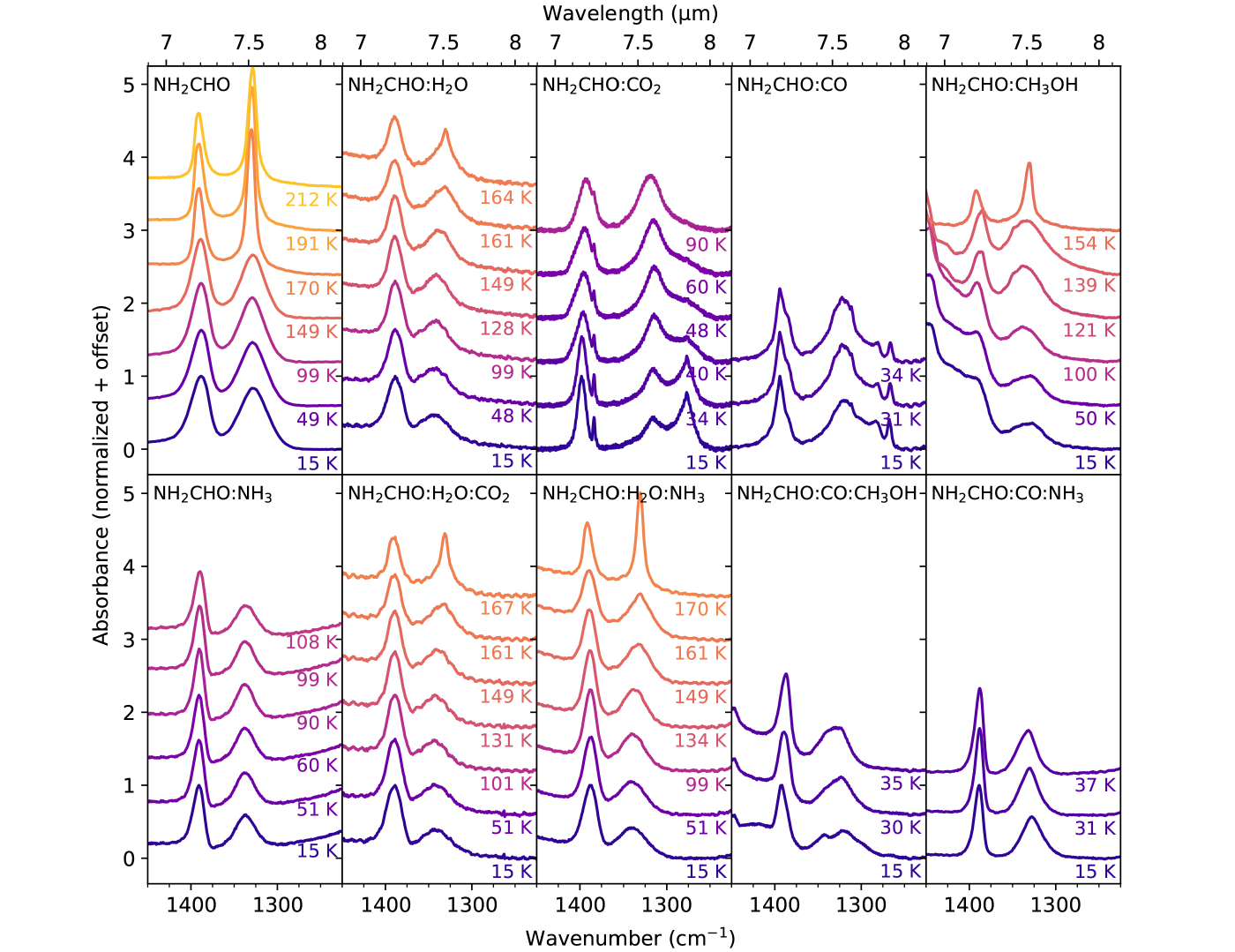}
\caption{CH bending and CN stretching bands of NH$_2$CHO in pure and mixed forms at selected temperatures.}
\label{fig:ch_bend_spectra}
\end{figure*}

\begin{figure*}
\centering
\includegraphics[width=0.95\textwidth]{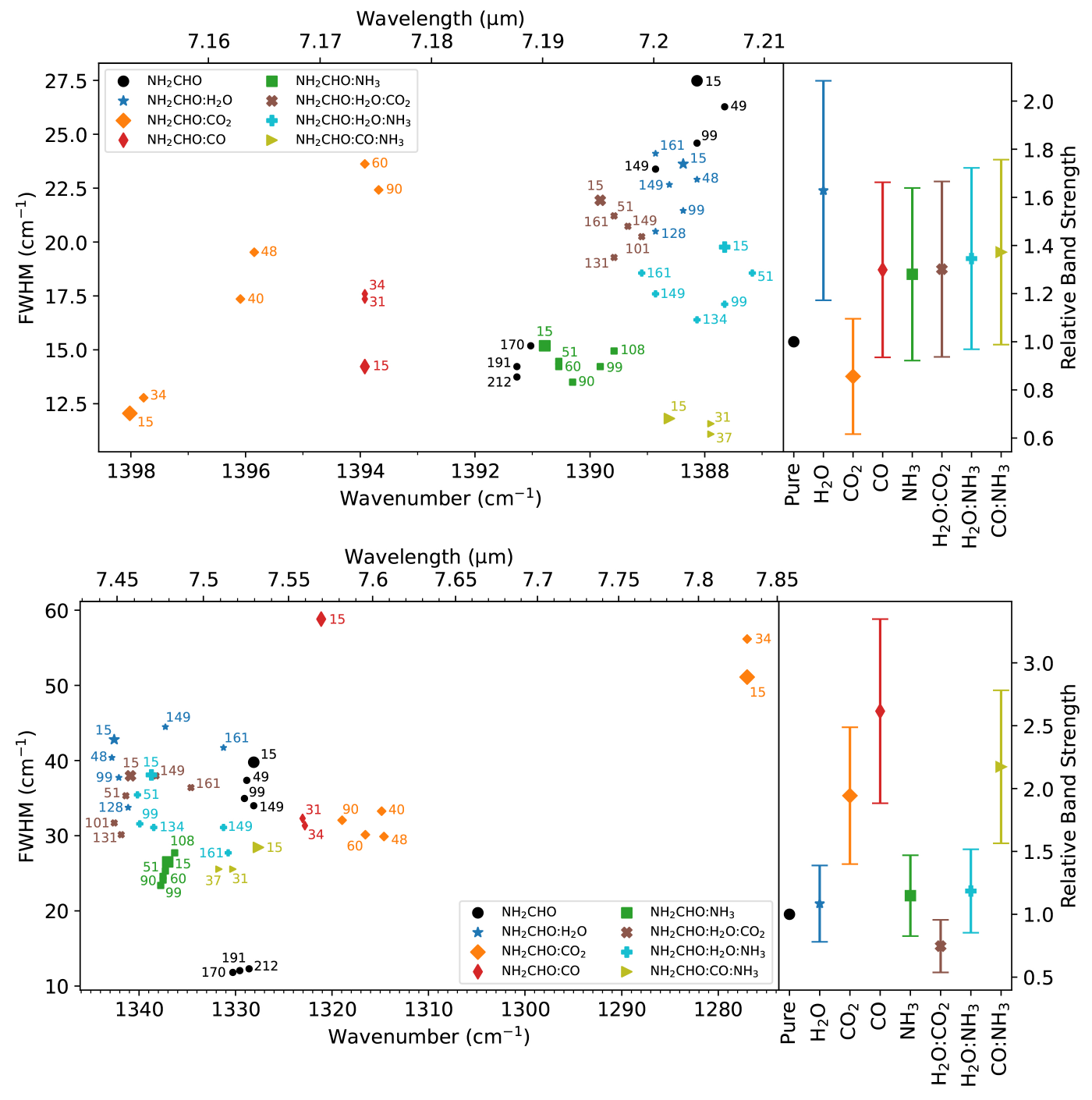}
\caption{Characterization of the CH bending and CN stretching bands of NH$_2$CHO. Left top: the peak position and FWHM of the formamide CH bend in pure and mixed forms at selected temperatures. Right top: the band strengths of the formamide CH bend in mixtures relative to the pure band strength (all at 15 K). Left bottom: the peak position and FWHM of the formamide CN stretch in pure and mixed forms at selected temperatures. Right bottom: the band strengths of the formamide CN stretch in mixtures relative to the pure band strength (all at 15 K).}
\label{fig:ch_bend_cn_stretch_boogert}
\end{figure*}

Figure~\ref{fig:co_stretch} shows how the profile of the C=O band (1700.3 cm$^{-1}$/5.881 $\mu$m) changes in different mixtures and temperatures and presents the peak positions and FWHMs of these spectra in a scatter plot. This type of scatter plot can help to narrow down the possible thermochemical environments of molecules identified in observations (see Section~\ref{astronomical_implications}). The right bottom plot in the figure shows the strengths of the band in the different mixtures at 15 K relative to the value of the band strength of pure formamide. The integrated regions used to calculate these band strengths also include the NH$_{2}$ scissoring mode, which presents as a weak, broad feature overlapping with the red shoulder of the C=O stretch (see Figure~\ref{fig:co_stretch}). The FWHM and relative band strengths of the formamide:NH$_{3}$ mixture are excluded from the bottom scatter plots in Figure~\ref{fig:co_stretch} and the tables in Appendix~\ref{appendix_a} due to the significant overlap of this band with ammonia's NH scissoring mode at 1624 cm$^{-1}$/6.16 $\mu$m. The NH$_{3}$ peak is small enough in the NH$_{3}$-containing tertiary mixtures relative to the formamide C=O stretch to extract reliable peak positions and FWHMs, but relative band strengths were not calculated. 

In pure amorphous formamide ($<$170 K), the C=O stretch appears as a single broad peak centered at 1704.2 cm$^{-1}$/5.868 $\mu$m. Generally, being in a mixture causes the feature to sharpen, most dramatically so in apolar mixtures in which CO or CO$_{2}$ are the dominant species. For example, the FWHM of the feature in formamide:CO$_{2}$ at 15 K is 51.1 cm$^{-1}$, over three times narrower than that in pure formamide. Also, in the CO, CO:CH$_{3}$OH, and crystalline CO$_{2}$ matrices, some peak splitting occurs before the formamide crystallization temperature is reached. Such sharpening and splitting is typical when a polar molecule is highly diluted in an apolar matrix and is caused by the polar molecule being isolated in the matrix as a monomer or dimer, unable to form the hydrogen bonds with other polar molecules that tend to broaden and blend vibrational features (e.g., \citealt{ehrenfreund1996infrared}). \citet{urso2017infrared} also previously observed the formamide peaks splitting due to monomer and dimer formation in their very dilute 1:40 formamide:CO mixture. In the polar mixtures, however, as hydrogen bonding with the matrix is still possible, the feature remains broad. The feature is the most blue-shifted in the binary CO and CO$_{2}$ mixtures, where its peak values are 1717.2 and 1703.7 cm$^{-1}$, respectively, in the 15 K ices, while in polar mixtures it tends to red-shift, with the most red-shifted peak position being that of the tertiary H$_{2}$O:CO$_{2}$ mixture, 1694.0 cm$^{-1}$. Despite containing a high fraction of apolar CO, the tertiary mixtures with CO:CH$_{3}$OH and CO:NH$_{3}$ have peak positions similar to the polar mixtures. The relative band strength of this formamide feature is $>$1 in all of the investigated matrices, with no observable trend related to polarity present in these values.

At formamide's crystalline phase transition temperature (170 K), the C=O peak blue-shifts and splits into multiple blended features. This is only observed in the pure formamide spectrum because all of the matrix molecules investigated here desorb below 170 K. An interesting trend to note is that, as the mixtures increase in temperature, the formamide C=O feature tends to broaden to have a FWHM value more similar to that of pure formamide. This trend can be easily identified in the scatter plot in Figure~\ref{fig:co_stretch}, where the scatter points of several of the mixtures move closer to the points of the pure amorphous spectrum as temperature increases. It is also particularly noticeable in Figure~\ref{fig:co_stretch} in the spectra of mixtures containing H$_{2}$O, which have peak position and FWHM values at high temperatures ($>$150 K) that are the close to those of the pure spectrum. Sudden broadening of the FWHM to a value closer to that of pure formamide also tends to occur at the matrix crystallization temperatures (for example, in the binary CO$_{2}$ mixture between $\sim$30 and 40 K and in the H$_{2}$O-containing mixtures between $\sim$130 and 150 K). These spectral changes indicate that formamide segregation is occurring in the matrix as the ice is heated and is particularly promoted when the ice undergoes a dramatic restructuring during matrix crystallization. The conclusion that solid-phase formamide diluted in a matrix is mobilized via heating is consistent with formamide thermal processing studies, in which formamide deposited on top of water ice diffused through the water during heating \citep{chaabouni2018thermal}.

\subsection{CH bending and CN stretching features ($\sim$1388 and 1328 cm$^{-1}$)}

The shape and position of the CH bend (1388.1 cm$^{-1}$/7.204 $\mu$m) does not vary much depending on chemical environment or temperature, with peak positions only ranging from 1398.0 - 1387.2 cm$^{-1}$ and FWHM values ranging from 11.1 - 27.5 cm$^{-1}$ in the mixtures investigated here (see Figures~\ref{fig:ch_bend_spectra} and \ref{fig:ch_bend_cn_stretch_boogert}). As in the C=O stretch band, the binary apolar mixtures with CO and CO$_{2}$ have the most blue-shifted and narrow peaks; however, a trend of the mixture band shifting during heating to peak position and FWHM values closer to those of the pure band is not as clear. The band strength of the CH bend increases in all of the mixtures (e.g., $\eta$=1.63 at 15 K in the formamide:H$_{2}$O mixture) except for the CO$_{2}$ mixture, in which the band strength decreases slightly ($\eta$=0.85 at 15 K).

The CN stretching band (1328.1 cm$^{-1}$/7.529 $\mu$m) varies much more dramatically across different mixtures and temperatures (see Figures~\ref{fig:ch_bend_spectra} and \ref{fig:ch_bend_cn_stretch_boogert}), particularly in the binary apolar mixtures, in which it red-shifts by up to $\sim$50 cm$^{-1}$ and splits into multiple convolved features. In the formamide:CO$_{2}$ spectrum, two peaks are present at 15 K at 1316.8 and 1277.0 cm$^{-1}$, with the peak at 1277.0 cm$^{-1}$ having a greater intensity until 40 K, at which point the intensity of the 1316.8 cm$^{-1}$ peak increases and that of the 1277.0 cm$^{-1}$ peak decreases. The 1277.0 cm$^{-1}$ peak intensity then continues to decrease during heating until CO$_{2}$ sublimates at 90 K (see Figure~\ref{fig:ch_bend_spectra}). This trend is indicative of the 1277.0 cm$^{-1}$ peak belonging to the formamide monomer and the 1316.8 cm$^{-1}$ peak belonging to the formamide dimer, as it would be expected for the monomer peak to decrease and the dimer peak to increase if segregation occurs during heating, especially during a major ice structure rearrangement like matrix crystallization, which occurs for CO$_{2}$ at 40 K. Such assignments are consistent with the assignments in \citet{mardyukov2007formamide}, who observed the formamide monomer and dimer in a xenon matrix at 1267.2 and 1305.4 cm$^{-1}$, respectively, and supported their assignments with computations. The peak in the formamide:CO spectrum also has a red component that appears to decrease in intensity during heating, but the monomer and dimer peaks are not as clearly distinguishable as more than two peaks appear to be overlapping in that spectrum. In the mixtures containing other polar molecules, the band is generally blue-shifted, broadened, and decreases in intensity relative to the CH bend. The relative strength of the band is close to 1 in most of the characterized polar mixtures, except for the H$_{2}$O:CO$_{2}$ mixture, which has a relative band strength of 0.75 at 15 K. In contrast, the relative band strength is closer to 2 in all of the primarily apolar mixtures.

While the CN stretch clearly has more potential than the CH bend as a diagnostic of the chemical environment of formamide, it is also much broader and less intense in most of the mixture spectra than in the pure spectra. This diminishes the ability to identify this band in a spectral region where several other astronomically relevant COMs also have features (see Section~\ref{astronomical_implications}).

\section{Astronomical implications}
\label{astronomical_implications}

The ability of formamide to form via both atom addition and energetic processing in a variety of ices containing C, H, N, and O means that its solid-state presence is plausible in many interstellar environments, ranging from dark interstellar clouds to protoplanetary disks. However, in order to securely detect it, an absorption with a clear peak position and profile that is distinguishable from other ice features in the same spectral region must be identified.

\begin{figure}[!ht]
\centering
\includegraphics[width=\linewidth]{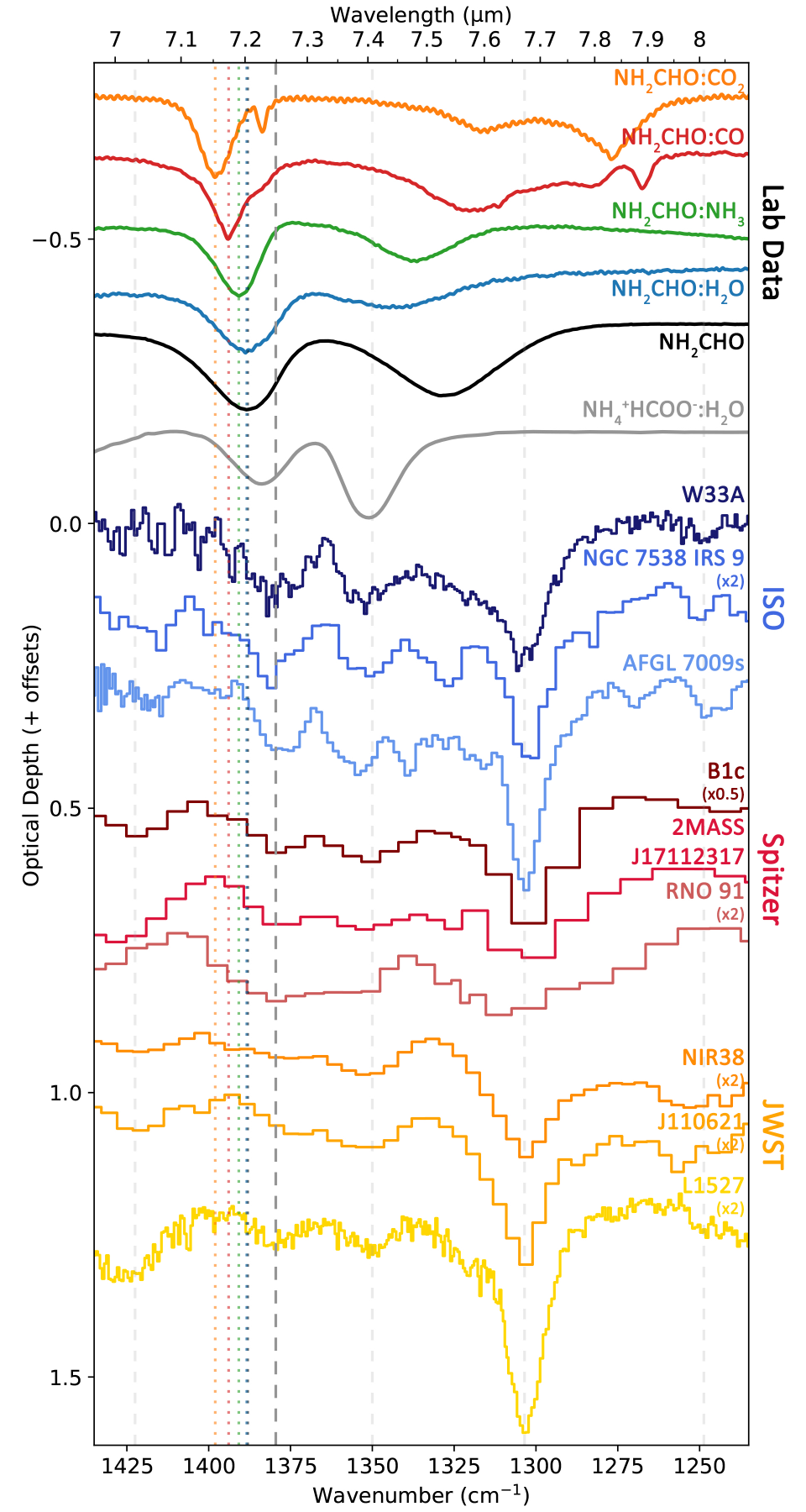}
\caption{Laboratory spectra of several formamide mixtures (all at 15 K) as well as an ammonium formate:water mixture from \citet{galvez2010ammonium} (at 150 K) versus local continuum-subtracted ISO spectra of three MYSOs (in blue), Spitzer spectra of three LYSOs (in red), and JWST spectra of both cold dense clouds in front of background stars and a LYSO (in gold). Optical depth scaling factors (if applied) are indicated via labels under the source names. Several of the band positions marked by gray dashed lines have been tentatively assigned to various COMs. The darkest gray dashed line indicates the peak position of the 7.24 $\mu$m band, and the colored dotted lines correspond to the peak positions of the formamide CH bend in the plotted formamide spectra. The wavelength calibration of the MIRI-LRS spectra, NIR38 and J110621, is still uncertain and was done for these spectra locally via the CH$_{4}$ band.}
\label{fig:com_rich_region}
\end{figure}

The C=O stretch is amorphous formamide's strongest and sharpest feature, but it overlaps with the blue wing of the strong and broad 6.0 $\mu$m feature present in most interstellar ice spectra. Water and ammonia, which have been securely identified in ices, as well as formic acid and formaldehyde, which have been tentatively identified, have features in this spectral region \citep{boogert2008c2d,boogert2015observations}. Additionally, many other carbonyl group-containing COMs that have been detected in the gas-phase and may be present in the solid state, like acetaldehyde, acetone, and methyl formate, also have strong absorptions in this wavelength region \citep{van2018infrared,rachid2020infrared,van2021infrared}. While this limits the potential of using formamide's C=O band as its primary means of identification, the band can still be used for performing fits spanning a wider wavelength region in combination with other bands.

The CH bend and the CN stretch are medium-strength features that lie in the "COM-rich region" of interstellar ice spectra between 7-8 $\mu$m \citep{boogert2008c2d}. This region, where many organic functional groups have absorptions, sits on the tail of the strong 6.85 $\mu$m band (whose assignment remains uncertain but likely contains absorptions by methanol and the ammonium cation, \citealt{boogert2008c2d,boogert2015observations}). The methane CH bending band at 7.68 $\mu$m is the most clearly and frequently observed ice band in this region \citep{oberg2008c2d}, but additional weaker features at 7.03, 7.24, 7.41, and 8.01 $\mu$m are also consistently observed toward some sources (Figure~\ref{fig:com_rich_region}). Candidate carriers suggested for some of these absorptions include species like formic acid, ethanol, acetaldehyde, the formate anion, and, potentially, formamide \citep{schutte1999weak,boogert2008c2d,van2018infrared}.

\begin{figure*}
\centering
\includegraphics[width=\linewidth]{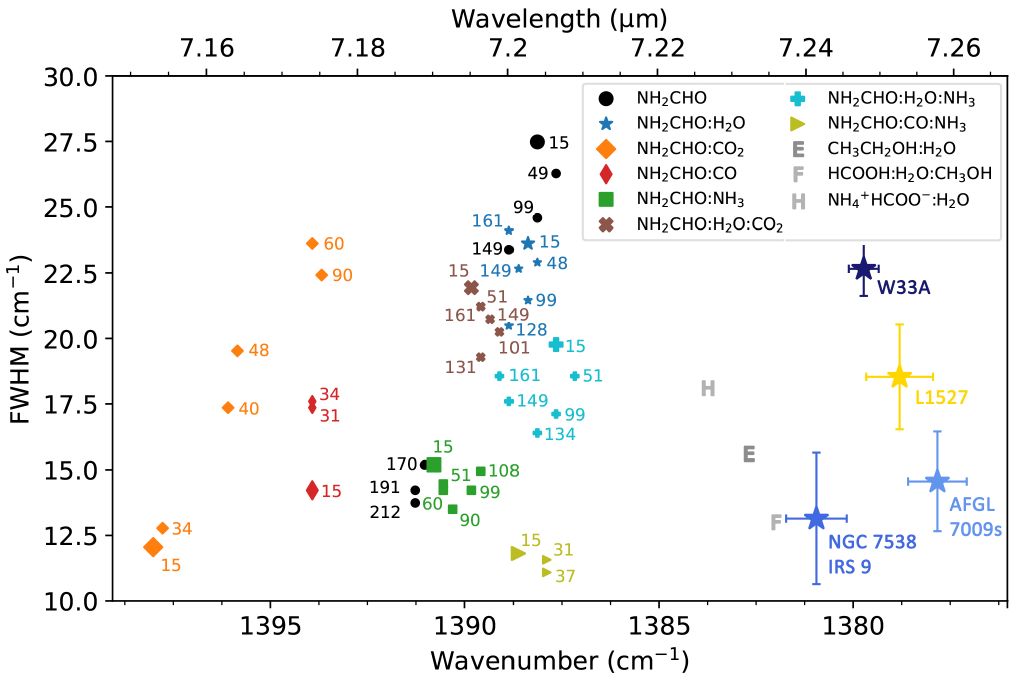}
\caption{Scatter plot containing all the same peak position and FWHM values of formamide's CH bend from Figure~\ref{fig:ch_bend_cn_stretch_boogert} along with the peak positions and FWHM values of the 7.24 $\mu$m peaks in the W33A, NGC 7538 IRS 9, AFGL 7009s, and L1527 spectra shown in Figure~\ref{fig:com_rich_region} and laboratory spectra of CH$_{3}$CH$_{2}$OH:H$_{2}$O \citep{van2018infrared}, HCOOH:H$_{2}$O:CH$_{3}$OH \citep{bisschop2007infrared} and NH$_{4}$$^{+}$HCOO$^{-}$:H$_{2}$O at 150 K \citep{galvez2010ammonium}.}
\label{fig:boogert_plot_obs}
\end{figure*}

As mentioned previously, \citet{schutte1999weak} tentatively assigned formamide as a plausible contributor to the 7.24 $\mu$m band in W33A using a formamide:H$_{2}$O spectrum and calculated a formamide ice upper limit of 2.1\% with respect to H$_{2}$O, although they pointed out that in their lab data, the formamide peak position was blue-shifted by 0.02 $\mu$m relative to the observed band, and that an assignment to the CH bend of formic acid (HCOOH) may be more appropriate. Ethanol (CH$_{3}$CH$_{2}$OH) and the formate anion (HCOO$^{-}$) have also been considered candidates for this band \citep{boogert2008c2d,oberg2011spitzer,van2018infrared,rocha2023james}. No distinct and consistently observed bands are located at the peak position of the formamide CN stretch at $\sim$7.5 $\mu$m. However, in mixtures (particularly those with polar components), the intensity and sharpness of this band weaken (relative to the intensity and sharpness of the CH bend). Such a profile change makes a distinction of the CN stretch from the continuum in this region less feasible if formamide is present at the low ice abundances expected for COMs, especially given that around this wavelength, many sources also show a broad and significant absorption commonly attributed to SO$_{2}$ ice \citep{boogert1997infrared,oberg2008c2d}. On the other hand, the CH bend remains strong and sharp in all of the mixtures investigated here. All of the other absorption features of formamide either have profiles that are too broad or weak, or overlap directly with the strongest absorptions of the major ice components (see Figure~\ref{fig:matrix_spectra}), and will therefore not be utilized in our hunt for formamide ice.

Thus, if formamide is indeed present in interstellar ices, the CH bend is likely its best tracer. We focus our subsequent analysis on the comparison of the formamide CH bend in mixtures to the observed 7.24 $\mu$m band in nine spectra collected toward a variety of sources by ISO, Spitzer, and the recently launched JWST (Figure~\ref{fig:com_rich_region}). The ISO (SWS) spectra include three massive young stellar objects (MYSOs), W33A, NGC 7538 IRS 9, and AFGL 7009s, and the Spitzer (IRS) spectra include three low-mass young stellar objects (LYSOs), B1c, 2MASS J17112317, and RNO 91. These archival spectra were selected due to their 7-8 $\mu$m regions having several deep and distinct features, indicating that they may be COM-rich, and because their profiles in this region slightly differ, demonstrating the variety of spectral features that have been observed here. In addition, three spectra recently collected by the JWST have been included: two pristine, high-extinction dark clouds toward background stars, NIR38 and J110621, observed with the Mid-InfraRed Instrument (MIRI) Low-Resolution Spectrometer (LRS) \citep{mcclure2023ice} in the ERS program Ice Age (program 1309, PI: M. McClure), and a Class 0/I low-mass protostar, L1527, observed with the MIRI Medium-Resolution Spectrometer (MRS) in the GTO program JOYS (program 1290, PI: E. F. van Dishoeck, \citealt{van2023diverse}). These are some of the first spectra ever collected of such low-flux sources. While the resolution of the ISO data is comparable to that of the JWST data, the resolution of the Spitzer data is significantly lower (R$\sim$60-100), limiting its use in the analysis of weak and narrow bands.

The 7.24 $\mu$m band is present to some extent in all of the sources, usually at an optical depth similar to the 7.41 $\mu$m band in the local continuum-subtracted spectra. The position and FWHM of the band were extracted from the spectra that have spectral resolutions high enough to clearly define the shape and position of the peak -- that is, the ISO-SWS and JWST MIRI-MRS MYSO spectra -- by fitting a Gaussian profile to the peak. Figure~\ref{fig:boogert_plot_obs} shows these observed peak positions and FWHMs (indicated with star shapes) in a scatter plot with the peak positions and FWHMs of the CH bend extracted from the laboratory spectra. The peak positions and FWHMs extracted from laboratory spectra of ethanol in a H$_{2}$O mixture \citep{van2018infrared}, formic acid in a H$_{2}$O:CH$_{3}$OH mixture \citep{bisschop2007infrared}, and ammonium formate in a H$_{2}$O mixture at 150 K \citep{galvez2010ammonium} are also included in this figure (indicated with the letters E, F, and H respectively) to enable a comparison between formamide and the other commonly proposed carriers. From this plot, it is evident that, while the polar mixtures have the band position and profile closest to the observations, they are all still too blue-shifted (by $\sim$7 cm$^{-1}$/0.04 $\mu$m) from the astronomical values for formamide to be the major carrier of this band. In contrast, ethanol, formic acid, and the formate anion in polar mixtures are much better candidates.

\begin{figure}[!ht]
\centering
\includegraphics[width=\linewidth]{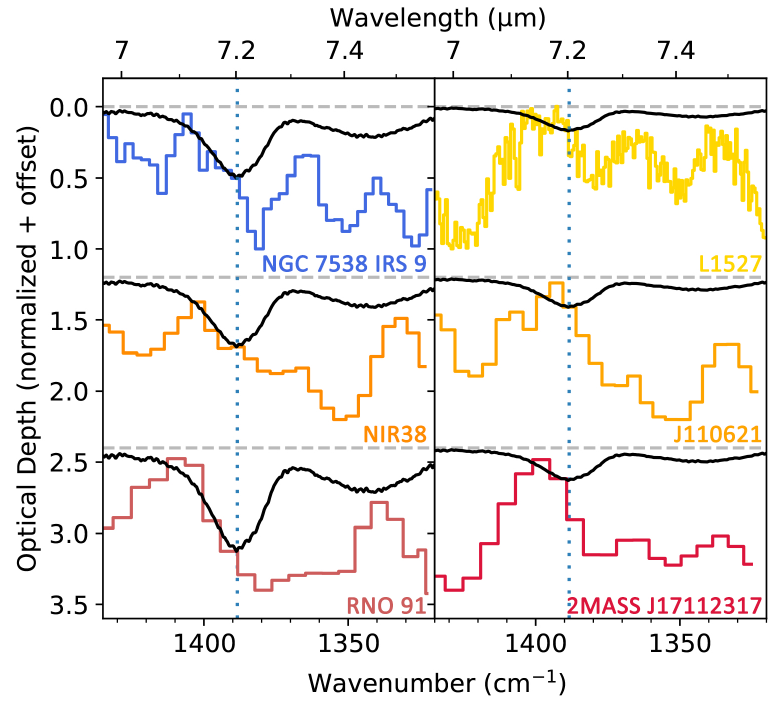}
\caption{Examples of fits of the CH bending band in the formamide:H$_{2}$O laboratory spectrum (in black, water contribution subtracted out) to the local-continuum subtracted observed spectra (in color) used to derive solid-state formamide upper limits. The peak position of the CH bend in the laboratory spectrum is marked with a dotted blue vertical line. These selected fits showcase the variation in the profile of the observed 7.24 $\mu$m band across the selected sources. In the observed spectra on the left, excess absorption on the blue wing of the observed 7.24 $\mu$m band allows for relatively high formamide upper limits ($\sim$$\le$1.5-5.1\% with respect to H$_{2}$O), while the observed spectra on the right lack such a wing, resulting in lower formamide upper limits ($\sim$$\le$0.35-0.68\% with respect to H$_{2}$O).}
\label{fig:fits_to_obs}
\end{figure}

It is still possible that formamide could be contributing to the blue wing of this band. However, to result in non-negligible upper limits, the formamide must be present in a matrix containing other polar molecules, as the band is far too blue-shifted in the purely apolar mixtures to contribute significantly to the observed absorption. Therefore, we derived upper limits of formamide by fitting the CH bend in the laboratory spectrum of the formamide:H$_{2}$O mixture at 15 K to the 7.24 $\mu$m band in the local continuum-subtracted observed spectra (see example fits in Figure~\ref{fig:fits_to_obs}). The water mixture was chosen for the fit for simplicity's sake and due to the fact that water is by far the most abundant interstellar ice component. The water contribution was subtracted out of the laboratory ice spectrum using a spectrum of pure water ice to ensure that absorption by the broad water bending band did not contribute to the calculated formamide upper limit. The band strength used to perform the upper limit calculation was 1.5$\times$10$^{-17}$ cm molec$^{-1}$, the band strength of the CH bend in pure formamide at 15 K (from Table~\ref{band-strength}) multiplied by the relative band strength of formamide in H$_{2}$O at 15 K (1.63, from Appendix~\ref{appendix_b}).

\begin{table}[h]
\caption{Upper limits of formamide with respect to H$_2$O compared to observed abundances of formamide in various objects.}
\begin{center}
\begin{tabular}{l c c}
\hline
        \multirow{2}{*}{Object} & \textit{N}(NH$_{2}$CHO) & \% H$_2$O \\
         & 10$^{17}$ molec cm$^{-2}$ & \\
        \hline
        \underline{MYSOs} & & \\
        W33A & $\le$2.8 & $\le$2.2 \\
        NGC 7538 IRS 9 & $\le$2.0 & $\le$3.1 \\
        AFGL 7009s & $\le$1.4 & $\le$1.2 \\
        & & \\
        \underline{LYSOs} & & \\
        B1c & $\le$2.8 & $\le$0.93 \\
        2MASS J17112317 & $\le$1.3 & $\le$0.68 \\
        RNO 91 & $\le$2.2 & $\le$5.1 \\
        L1527 & $\le$1.0 & $\le$0.50 \\
        & & \\
        \underline{Prestellar cores} & & \\
        NIR38 & $\le$1.0 & $\le$1.5 \\
        J110621 & $\le$0.47 & $\le$0.35 \\
    \hline
        \underline{LYSOs} & & \\
        B1c (gas) & - & $\sim$0.05$^a$ \\
        & & \\
        \underline{MYSOs} & & \\
        ALMAGAL average (gas) & - & $\sim$0.05$^b$ \\
        & & \\
        \underline{Comets} & & \\
        67P (dust and gas) & - & 0.004$^c$ \\
        Lemmon (gas) & - & 0.016$^d$ \\
        Lovejoy (gas) & - & 0.021$^d$ \\
        Hale-Bopp (gas) & - & 0.02$^e$ \\
    \hline
     \label{upper-lim-table}
\end{tabular}

\begin{tablenotes}
    \item[\emph{}]{$^{a}$ Value calculated via the beam dilution-corrected NH$_{2}$CHO/CH$_{3}$OH ratio from \citealt{nazari2023physical}, assuming a CH$_{3}$OH/H$_{2}$O ratio of $\sim$5\%.}
    \item[\emph{}]{$^{b}$ Value calculated via the beam dilution-corrected NH$_{2}$CHO/CH$_{3}$OH ratio of 22 MYSO sources from \citealt{nazari2023physical}, assuming a CH$_{3}$OH/H$_{2}$O ratio of $\sim$5\%.}
    \item[\emph{}]{$^{c}$ Value from \citealt{rubin2019elemental}.}
    \item[\emph{}]{$^{d}$ Value from \citealt{biver2014complex}.}
    \item[\emph{}]{$^{e}$ Value from \citealt{bockelee2000new}.}

\end{tablenotes}

\end{center}
\end{table}

When deriving upper limits, it is prudent to ensure that the laboratory spectrum fits to the observed spectrum across a wider wavelength range, as upper limits can be easily overestimated if only one band is considered. Subtracting out the contributions of other ices that absorb in the analyzed spectral region, if their abundances can be unambiguously determined from other spectral regions, also prevents further upper limit overestimations. Therefore, we ensured that the calculated upper limits in Table~\ref{upper-lim-table} do not result in a C=O stretch absorption that exceeds the observed optical depth of the $\sim$6 $\mu$m band in our selected objects. Prior to checking the C=O absorption in this region, the spectral contribution of water's OH bend $\sim$1655 cm$^{-1}$/6.04 $\mu$m was removed from the observed spectra by scaling a laboratory water spectrum at 15 K from \citet{oberg2007effects}, so that the water column density of the scaled spectrum was the same as what was previously determined for these objects, and then performing a subtraction. (For the ISO and Spitzer data, the water column densities from \citet{boogert2008c2d} were used for scaling; for the JWST MIRI-LRS data, the water column densities from \citet{mcclure2023ice} were used. For the JWST MIRI-MRS spectrum (L1527), the water column density was determined by first subtracting the silicate contribution by fitting the GCS3 spectrum to the 10 $\mu$m silicate band and then fitting the laboratory water spectrum from \citealt{oberg2007effects} to the water libration band.)

The resulting upper limits of solid-state formamide, presented in column densities as well as with respect to the abundance of water in each source, are presented in Table~\ref{upper-lim-table}. These upper limits (ranging from 0.35-5.1\% with respect to H$_{2}$O) are all at least an order of magnitude greater than (but consistent with) the observed gas-phase formamide abundances in three comets (0.016-0.021\% with respect to H$_{2}$O) as well as the average beam dilution-corrected abundance of 22 MYSOs from the ALMAGAL survey ($\sim$0.05\% with respect to H$_{2}$O, assuming a CH$_{3}$OH/H$_{2}$O ratio of $\sim$5\%). As a beam dilution-corrected gas-phase formamide abundance has also been obtained for the LYSO B1c ($\sim$0.05\%), one of the sources investigated here, it can be directly compared to our solid-state formamide upper limit derived from the object's low-resolution Spitzer data. While our upper limit ($\le$0.93\%) is consistent with this gas-phase abundance, it is an order of magnitude greater. We expect the precision of this upper limit to be further refined by future high-resolution observations of B1c, planned to be observed by MIRI-MRS in the JOYS program.

A formamide upper limit of 2.1\% with respect to H$_{2}$O was previously derived for W33A in \citet{schutte1999weak} by assuming that the entire 7.24 $\mu$m band consisted of formamide and using a band strength of 3.2 $\times$ 10$^{-18}$ cm molec$^{-1}$ attributed to \citet{wexler1967integrated}, where it is unclear for what phase of formamide this band strength was derived. Despite our very different approaches, we have fortuitously arrived at nearly the same upper limit value for W33A (2.2\%).

In the higher resolution observational data of MYSOs explored here, the lack of a formamide CH bending feature distinct from other COM absorptions prevents a secure formamide ice detection. However, it is clear from the example upper limit fits shown in Figure~\ref{fig:fits_to_obs} that the profile of the 7.24 $\mu$m feature is not uniform across different sources, and several sources, such as NGC 7538 IRS 9, NIR38, and RNO 91, may have a blue wing on this band that spectrally overlaps with the CH bend of formamide. Therefore, it is possible that a more distinct absorption at the expected 7.20 $\mu$m will emerge more clearly in sources targeted by future JWST MIRI-MRS observations. The first ice spectra arriving now from MIRI-MRS illuminate a promising future. In the spectrum of the LYSO IRAS 15398-3359 acquired by the JWST CORINOS program (program 2151, PI: Y. -L. Yang, \citealt{yang2022corinos}), the COM features between 7-8 $\mu$m previously detected barely above 3$\sigma$ levels in the spectra in Figure~\ref{fig:com_rich_region} are beautifully resolved (although a distinct absorption centered at 7.20 $\mu$m is not present). More sources known to have strong COM absorptions in this spectral region have been specifically targeted by the JOYS program as well as the JWST proposals "It's COMplicated" (program 1854, PI: M. McClure, \citealt{mcclure2021s}) and "Ice chemical complexity toward the Ophiuchus molecular cloud" (program 1959, PI: W. R. M. Rocha, \citealt{rocha2021ice}), and these sources are scheduled to be observed throughout the remainder of this year. As demonstrated by the examples of spectral analysis of ices in this section, the laboratory spectra from this work can serve as a toolkit for formamide identification in such ice observations.

\section{Conclusions}

In an effort to facilitate the hunt for formamide in interstellar ices, laboratory spectra of pure formamide and formamide in various astronomically relevant ice mixtures ranging from temperatures of 15 - 212 K have been collected and made freely available to the astronomical community on the Leiden Ice Database for Astrochemistry (LIDA). The band strengths at 15 K for all pure formamide features between 4000 - 500 cm$^{-1}$/2.5 - 20 $\mu$m are presented, and the peak positions, FWHMs, and relative apparent band strengths of the three bands identified as the most promising for future formamide detection were extracted from the pure and mixed formamide spectra. These spectra and extracted data were used to assess present and future detectability of ices in various interstellar objects. The primary conclusions drawn from this work are as follows:

\begin{enumerate}

\item Out of the eight formamide features in the investigated IR spectral region, the C=O stretch (1700.9 cm$^{-1}$/5.881 $\mu$m), the CH bend (1388.3 cm$^{-1}$/7.203 $\mu$m), and the CN stretch (1328.0 cm$^{-1}$/7.530 $\mu$m) are likely to be the most useful for future formamide identification due to their strength, sharp profile, and low overlap with the strongest features of the major ice components, with the CH bending feature being the most promising. The NH$_{2}$ stretching features (3371.2 cm$^{-1}$/2.966 $\mu$m and 3176.4 cm$^{-1}$/3.148 $\mu$m) and the NH$_{2}$ wagging and twisting features (689.2 cm$^{-1}$/14.510 $\mu$m and 634.0 cm$^{-1}$/15.773 $\mu$m) directly overlap with strong water absorptions, while the CH stretch (2881.9 cm$^{-1}$/3.470 $\mu$m), the CH bend overtone (2797.7 cm$^{-1}$/3.574 $\mu$m), and the convolved NH$_{2}$ rock and CH out-of-plane deformation (1108.1 cm$^{-1}$/9.024 $\mu$m and 1056.1 cm$^{-1}$/9.469 $\mu$m) have both low band strengths and direct overlap with methanol absorptions, making them less suitable for formamide identification.

\item In the mixtures investigated here, the CN stretch is the most affected by ice composition -- its peak position varies by up to $\sim$68 cm$^{-1}$ and its FWHM by up to $\sim$50 cm$^{-1}$ across the mixtures investigated here, with peak splitting observed in the apolar mixtures. The C=O stretch can also change significantly, depending on the matrix, by up to $\sim$27 cm$^{-1}$ in peak position and up to $\sim$40 cm$^{-1}$ in FWHM, although peak splitting in the apolar mixtures is not as prominent as in the CN stretch. The CH bend is relatively unaffected by ice composition, with its peak position and FWHM only varying by $\sim$11 cm$^{-1}$ and $\sim$15 cm$^{-1}$, respectively, across the different mixtures. Relative to the pure spectrum, the band strength of the C=O stretch increases in all of the investigated mixtures. The CH bend band strength also increases in all of the mixtures except the binary CO$_{2}$ mixture, while a significant increase in the band strength of the CN stretch is only observed in the mixtures dominated by an apolar component.

\item Although the polar formamide mixtures provide the closest match to the 7.24 $\mu$m band observed toward nine lines of sight (including dense clouds, LYSOs, and MYSOs) with three different space telescopes (ISO, Spitzer, and JWST), none provide a convincing fit, with all having their CH bend peak position approximately 7 cm$^{-1}$/0.04 $\mu$m too far to the blue from the clearly observed band at 1381 cm$^{-1}$/7.24 $\mu$m. Instead, formic acid and ethanol mixtures containing H$_{2}$O provide a better fit. However, this does not exclude the possibility of formamide being present in these ices. The calculated formamide upper limits in these objects range from 0.35-5.1\% with respect to H$_{2}$O, which are consistent with gas-phase abundances of formamide in several LYSOs, MYSOs, and comets. The upper limit value derived for W33A, 2.2\% with respect to H$_{2}$O, is fortuitously in agreement with that derived by \citet{schutte1999weak}.

\end{enumerate}

While a more secure formamide detection is not possible with the telescopic data explored in this work, the first ice observations arriving from JWST demonstrate an unprecedented sensitivity and spectral resolution that will enable us in the near future to broaden the search for formamide ice in both objects previously observed by Spitzer, whose analysis is limited by low spectral resolution, as well as newly observed objects that were too dim to be observed by Spitzer or ISO.

\begin{acknowledgements}
  This work is supported by funding from the European Research Council (ERC) under the European Union’s Horizon 2020 research and innovation programme (grant agreement No. 101019751 MOLDISK), the Netherlands Research School for Astronomy (NOVA), and the Danish National Research Foundation through the Center of Excellence "InterCat" (Grant agreement no.: DNRF150). The authors acknowledge the Ice Age (program 1309, PI: M. McClure) and JOYS (program 1290, PI: E. F. van Dishoeck) observing programs for the JWST astronomical data used in this work. KS acknowledges Thanja Lamberts and Pooneh Nazari for helpful discussions about the formamide formation mechanism and Sergio Ioppolo for helpful discussions about the QMS calibration methodology.
\end{acknowledgements}


\bibliographystyle{aa}
\bibliography{biblio}

\begin{appendix}

\section{Peak positions and FWHMs of formamide in pure and mixed ices}
\label{appendix_a}

This appendix contains the peak positions and FWHMs of the formamide features selected for complete IR characterization in this work. The values are listed for the formamide features in pure ice as well as in mixtures containing H$_{2}$O, CO$_{2}$, CO, CH$_{3}$OH, and NH$_{3}$. The peak position is the wavelength at which the absorption reaches its maximum, and the FWHM is the width of the peak between the half-maximum values on each side. A Savitzky-Golay filter with a second-order polynomial was applied to many of the mixture spectra here before extraction of the peak position and FWHM to eliminate shifts in these values caused by noise. The smoothing windows used ranged from 10-100 depending on level of noise present in each spectrum, and care was taken that these smoothing windows did not warp the shape of any features. Values were extracted until the temperatures at which the major matrix component desorbed were reached.

For formamide features in mixtures where there is direct overlap with weaker matrix component bands (e.g., the C=O stretch in the NH$_{2}$CHO:H$_{2}$O mixture), the spectrum of the matrix component without formamide collected using identical experimental parameters at the corresponding temperature was scaled to the formamide mixture spectrum via a feature without overlap with formamide features and subtracted prior to peak position and FWHM extraction. These cases are denoted with a $^{M}$. For formamide features in mixtures where the formamide features lie on the tails of bands or on very wide bands without sharp features (e.g., the CH bend and CN stretch in the NH$_{2}$CHO:NH$_{3}$ mixture), a second-order polynomial was used to perform a local continuum subtraction. These cases are denoted with a $^{P}$. For formamide features in mixtures where overlap with a strong matrix component band was very substantial and difficult to reliably subtract (e.g., the C=O stretch in the NH$_{2}$CHO:NH$_{3}$ mixture), only peak positions are given. These cases are denoted with a $^{N}$. For formamide features that contain multiple peaks, all peak positions are given, and the FWHM of the strongest peak is given. However, if a weaker peak maximum occurs within the two half maximum values of the stronger peak (e.g., the CN stretch in the NH$_{2}$CHO:CO 15 K mixture), it is included in the FWHM. These cases are denoted with a $^{B}$.

\begin{table}[h]
\caption[]{Peak positions and FWHMs of the C=O stretch and NH$_{2}$ scissor bands of pure formamide and formamide in various binary mixtures at temperatures from 15-212 K.}
\fontsize{9}{10}\selectfont 
\begin{center}
\begin{tabular}{|c|c|cc|cc|}
\hline
         \multirow{2}{*}{Matrix} & \multirow{2}{*}{T (K)} & \multicolumn{2}{c|}{Peak} & \multicolumn{2}{c|}{FWHM} \\
           &  & (cm$^{-1}$) & ($\mu$m) & (cm$^{-1}$) & ($\mu$m)\\
        \hline 
         \multirow{17}{*}{Pure} & 15 & 1700.3 & 5.881 & 58.8 & 0.203 \\
         & & 1630.4 & 6.133 & - & - \\
         & 49 & 1700.3 & 5.881 & 58.1 & 0.201 \\
         & & 1632.6 & 6.125 & - & - \\
         & 99 & 1700.3 & 5.881 & 56.6 & 0.196 \\
         & & 1631.1 & 6.131 & - & - \\
         & 149 & 1699.8 & 5.883 & 55.4 & 0.192 \\
         & & 1629.4 & 6.137 & - & - \\
         & 170 & 1704.2 & 5.868 & 44.1 & 0.149 \\
         & & 1749.2 & 5.717 & - & - \\
         & & 1647.0 & 6.072 & - & - \\
         & 191 & 1703.4 & 5.870 & 44.8 & 0.152 \\
         & & 1750.0 & 5.714 & - & - \\
         & & 1644.6 & 6.080 & - & - \\
         & 212 & 1704.2 & 5.868 & 45.3 & 0.154 \\
         & & 1750.4 & 5.713 & - & - \\
         & & 1642.2 & 6.089 & - & - \\
         \hline
        \multirow{6}{*}{H$_{2}$O$^{M}$} & 15 & 1701.0 & 5.879 & 36.9 & 0.128 \\
         & 49 & 1701.0 & 5.879 & 37.4 & 0.129 \\
         & 99 & 1701.0 & 5.879 & 37.6 & 0.130 \\
         & 129 & 1701.3 & 5.878 & 38.1 & 0.132 \\
         & 149 & 1702.2 & 5.875 & 47.5 & 0.164 \\
         & 161 & 1703.0 & 5.872 & 54.0 & 0.187 \\
         \hline
        \multirow{12}{*}{CO$_{2}$} & 15 & 1703.7 & 5.870 & 17.6 & 0.061 \\
         & & 1591.6 & 6.283 & - & - \\
         & 34 & 1704.2 & 5.868 & 18.8 & 0.065 \\
         & & 1591.6 & 6.283 & - & - \\
         & 40 & 1709.5 & 5.850 & 27.2 & 0.093 \\
         & & 1591.6 & 6.283 & - & - \\
         & 49 & 1711.2 & 5.844 & 27.0 & 0.093 \\
         & & 1615.2 & 6.191 & - & - \\
         & 60 & 1710.9 & 5.845 & 29.9 & 0.103 \\
         & & 1614.0 & 6.196 & - & - \\
         & 90 & 1709.5 & 5.850 & 38.8 & 0.134 \\
         & & 1613.8 & 6.197 & - & - \\
         \hline
         \multirow{6}{*}{CO} & 15 & 1717.2 & 5.824 & 34.7 & 0.119 \\
         & & 1588.9 & 6.294 & - & - \\
         & 31 & 1717.2 & 5.824 & 36.4 & 0.125 \\
         & & 1588.7 & 6.294 & - & - \\
         & 34 & 1716.9 & 5.824 & 36.6 & 0.126 \\
         & & 1588.7 & 6.294 & - & - \\
         \hline
         \multirow{12}{*}{CH$_{3}$OH$^{M}$} & 15 & 1696.7 & 5.894 & 25.8 & 0.090 \\
         & & 1623.5 & 6.159 & - & - \\
         & 50 & 1696.9 & 5.893 & 25.8 & 0.090 \\
         & & 1623.3 & 6.160 & - & - \\
         & 100 & 1697.4 & 5.891 & 29.9 & 0.104 \\
         & & 1621.4 & 6.168 & - & - \\
         & 121 & 1700.1 & 5.882 & 32.1 & 0.111 \\
         & & 1621.4 & 6.168 & - & - \\
         & 139 & 1700.8 & 5.880 & 40.0 & 0.139 \\
         & & 1621.4 & 6.168 & - & - \\
         & 154 & 1720.1 & 5.814 & 46.5 & 0.156 \\
         & & 1646.9 & 6.072 & - & - \\
         \hline
         \multirow{6}{*}{NH$_{3}$$^{N}$} & 15 & 1698.4 & 5.888 & - & - \\
         & 51 & 1698.1 & 5.889 & - & - \\
         & 60 & 1697.9 & 5.890 & - & - \\
         & 90 & 1696.9 & 5.893 & - & - \\
         & 99 & 1697.7 & 5.890 & - & - \\
         & 108 & 1698.9 & 5.886 & - & - \\
        \hline
        
\end{tabular}
\end{center}
\label{co_peaks_fwhm_1}
\end{table}

\begin{table}[h]
\caption[]{Peak positions and FWHMs of the C=O stretch and NH$_{2}$ scissor bands of formamide in various tertiary mixtures at temperatures from 15-161 K.}
\fontsize{9}{10}\selectfont 
\begin{center}
\begin{tabular}{|c|c|cc|cc|}
\hline
         \multirow{2}{*}{Matrix} & \multirow{2}{*}{T (K)} & \multicolumn{2}{c|}{Peak} & \multicolumn{2}{c|}{FWHM} \\
           &  & (cm$^{-1}$) & ($\mu$m) & (cm$^{-1}$) & ($\mu$m)\\
        \hline 
        \multirow{6}{*}{H$_{2}$O:CO$_{2}$$^{M}$} & 15 & 1694.0 & 5.903 & 36.4 & 0.127 \\
         & 51 & 1693.6 & 5.905 & 37.6 & 0.131 \\
         & 101 & 1696.4 & 5.895 & 38.6 & 0.134 \\
         & 131 & 1697.9 & 5.890 & 38.3 & 0.133 \\
         & 149 & 1699.1 & 5.885 & 43.1 & 0.150 \\
         & 161 & 1700.8 & 5.880 & 48.9 & 0.169 \\
        \hline
        \multirow{6}{*}{H$_{2}$O:NH$_{3}$$^{M}$} & 15 & 1698.1 & 5.889 & 36.6 & 0.127 \\
         & 51 & 1698.1 & 5.889 & 37.1 & 0.129 \\
         & 99 & 1697.9 & 5.890 & 35.2 & 0.122 \\
         & 134 & 1698.1 & 5.889 & 37.6 & 0.131 \\
         & 149 & 1697.9 & 5.890 & 46.8 & 0.162 \\
         & 161 & 1696.4 & 5.895 & 59.3 & 0.206 \\
        \hline
        \multirow{6}{*}{CO:CH$_{3}$OH$^{P}$} & 15 & 1695.5 & 5.898 & 27.2 & 0.095 \\
        & & 1615.2 & 6.191 & - & - \\
         & 30 & 1696.2 & 5.896 & 28.9 & 0.100 \\
         & & 1615.2 & 6.191 & - & - \\
         & 35 & 1700.3 & 5.881 & 28.4 & 0.099 \\
         & & 1617.1 & 6.184 & - & - \\
        \hline
        \multirow{3}{*}{CO:NH$_{3}$$^{M}$} & 15 & 1698.1 & 5.889 & 25.3 & 0.088 \\
         & 30 & 1697.9 & 5.890 & 26.0 & 0.090 \\
         & 37 & 1698.6 & 5.887 & 27.7 & 0.096 \\
        \hline
        
\end{tabular}
\end{center}
\label{co_peaks_fwhm_2}
\end{table}

\begin{table}[h]
\caption[]{Peak positions and FWHMs of the CH bend band of pure formamide and formamide in various binary and tertiary mixtures at temperatures from 15-212 K.}
\fontsize{9}{10}\selectfont 
\begin{center}
\begin{tabular}{|c|c|cc|cc|}
\hline
         \multirow{2}{*}{Matrix} & \multirow{2}{*}{T (K)} & \multicolumn{2}{c|}{Peak} & \multicolumn{2}{c|}{FWHM} \\
           &  & (cm$^{-1}$) & ($\mu$m) & (cm$^{-1}$) & ($\mu$m)\\
        \hline 
         \multirow{7}{*}{Pure} & 15 & 1388.1 & 7.204 & 27.5 & 0.142 \\
         &49 & 1387.7 & 7.206 & 26.3 & 0.136 \\
         &99 & 1388.1 & 7.204 & 24.6 & 0.127 \\
         &149 & 1388.9 & 7.200 & 23.4 & 0.121 \\
         &170 & 1391.0 & 7.189 & 15.2 & 0.079 \\
         &191 & 1391.3 & 7.188 & 14.2 & 0.074 \\
         &212 & 1391.3 & 7.188 & 13.7 & 0.071 \\
        \hline
        \multirow{6}{*}{H$_{2}$O$^{M}$} & 15 & 1388.4 & 7.203 & 23.6 & 0.122 \\
         & 49 & 1388.1 & 7.204 & 22.9 & 0.119 \\
         & 99 & 1388.4 & 7.203 & 21.5 & 0.111 \\
         & 129 & 1388.9 & 7.200 & 20.5 & 0.106 \\
         & 149 & 1388.6 & 7.201 & 22.7 & 0.117 \\
         & 161 & 1388.9 & 7.200 & 24.1 & 0.125 \\
        \hline
        \multirow{12}{*}{CO$_{2}$} & 15 & 1398.0 & 7.153 & 12.1 & 0.062 \\
         & & 1383.7 & 7.227 & - & - \\
         & 34 & 1397.8 & 7.154 & 12.8 & 0.065 \\
         & & 1383.7 & 7.227 & - & - \\
         & 40 & 1396.1 & 7.163 & 17.4 & 0.089 \\
         & & 1383.9 & 7.226 & - & - \\
         & 49 & 1395.8 & 7.164 & 19.5 & 0.100 \\
         & & 1383.9 & 7.226 & - & - \\
         & 60 & 1393.9 & 7.174 & 23.6 & 0.122 \\
         & & 1383.9 & 7.226 & - & - \\
         & 90 & 1393.7 & 7.175 & 22.4 & 0.116 \\
         & & 1384.4 & 7.223 & - & - \\
         \hline
         \multirow{3}{*}{CO} & 15 & 1393.9 & 7.174 & 14.2 & 0.073 \\
         & 31 & 1393.9 & 7.174 & 17.4 & 0.090 \\
         & 34 & 1393.9 & 7.174 & 17.6 & 0.091 \\
         \hline
         \multirow{6}{*}{NH$_{3}$$^{P}$} & 15 & 1390.8 & 7.190 & 15.2 & 0.078 \\
         & 51 & 1390.5 & 7.191 & 14.5 & 0.075 \\
         & 60 & 1390.5 & 7.191 & 14.2 & 0.073 \\
         & 90 & 1390.3 & 7.193 & 13.5 & 0.070 \\
         & 99 & 1389.8 & 7.195 & 14.2 & 0.074 \\
         & 108 & 1389.6 & 7.196 & 14.9 & 0.077 \\
        \hline
        \multirow{6}{*}{H$_{2}$O:CO$_{2}$$^{M}$} & 15 & 1389.9 & 7.195 & 21.9 & 0.114 \\
         & 51 & 1389.6 & 7.196 & 21.2 & 0.110 \\
         & 101 & 1389.1 & 7.199 & 20.2 & 0.105 \\
         & 131 & 1389.6 & 7.196 & 19.3 & 0.100 \\
         & 149 & 1389.3 & 7.198 & 20.7 & 0.107 \\
         & 161 & 1389.6 & 7.196 & 21.2 & 0.110 \\
        \hline
        \multirow{6}{*}{H$_{2}$O:NH$_{3}$$^{P}$} & 15 & 1387.7 & 7.206 & 19.8 & 0.103 \\
         & 51 & 1387.2 & 7.209 & 18.6 & 0.096 \\
         & 99 & 1387.7 & 7.206 & 17.1 & 0.089 \\
         & 134 & 1388.1 & 7.204 & 16.4 & 0.085 \\
         & 149 & 1388.9 & 7.200 & 17.6 & 0.091 \\
         & 161 & 1389.1 & 7.199 & 18.6 & 0.096 \\
        \hline
        \multirow{3}{*}{CO:NH$_{3}$$^{M}$} & 15 & 1388.6 & 7.201 & 11.8 & 0.061 \\
         & 30 & 1387.9 & 7.205 & 11.1 & 0.058 \\
         & 37 & 1387.9 & 7.205 & 11.6 & 0.060 \\
        \hline
        
\end{tabular}
\end{center}
\label{ch_peaks_fwhm}
\end{table}

\begin{table}[h]
\caption[]{Peak positions and FWHMs of the CN stretch band of pure formamide and formamide in various binary and tertiary mixtures at temperatures from 15-212 K.}
\fontsize{9}{10}\selectfont 
\begin{center}
\begin{tabular}{|c|c|cc|cc|}
\hline
         \multirow{2}{*}{Matrix} & \multirow{2}{*}{T (K)} & \multicolumn{2}{c|}{Peak} & \multicolumn{2}{c|}{FWHM} \\
           &  & (cm$^{-1}$) & ($\mu$m) & (cm$^{-1}$) & ($\mu$m)\\
        \hline 
         \multirow{7}{*}{Pure} & 15 & 1328.1 & 7.529 & 39.8 & 1.226 \\
         & 49 & 1328.8 & 7.525 & 37.4 & 0.212 \\
         & 99 & 1329.1 & 7.524 & 35.0 & 0.198 \\
         & 149 & 1328.1 & 7.529 & 34.0 & 0.192 \\
         & 170 & 1330.3 & 7.517 & 11.8 & 0.067 \\
         & 191 & 1329.6 & 7.521 & 12.1 & 0.068 \\
         & 212 & 1328.6 & 7.527 & 12.3 & 0.070 \\
        \hline
        \multirow{6}{*}{H$_{2}$O$^{M}$} & 15 & 1342.6 & 7.448 & 42.8 & 0.237 \\
         & 49 & 1342.8 & 7.447 & 40.4 & 0.223 \\
         & 99 & 1342.1 & 7.451 & 37.7 & 0.209 \\
         & 129 & 1341.1 & 7.456 & 33.7 & 0.188 \\
         & 149 & 1337.3 & 7.478 & 44.5 & 0.248 \\
         & 161 & 1331.2 & 7.512 & 41.7 & 0.234 \\
        \hline
        \multirow{8}{*}{CO$_{2}$} & 15 & 1277.0 & 7.831 & 51.1 & 0.305 \\
         & & 1316.8$^{B}$ & 7.594 & - & - \\
         & 34 & 1277.0 & 7.831 & 56.2 & 0.334 \\
         & & 1316.8$^{B}$ & 7.594 & - & - \\
         & 40 & 1314.9 & 7.605 & 33.3 & 0.194 \\
         & 49 & 1314.6 & 7.607 & 29.9 & 0.173 \\
         & 60 & 1316.5 & 7.596 & 30.1 & 0.174 \\
         & 90 & 1319.0 & 7.582 & 32.1 & 0.184 \\
         \hline
         \multirow{9}{*}{CO} & 15 & 1321.1 & 7.569 & 58.8 & 0.344 \\
         & & 1283.3$^{B}$ & 7.793 & - & - \\
         & & 1267.6 & 7.889 & - & - \\
         & 31 & 1323.0 & 7.558 & 32.3 & 0.185 \\
         & & 1280.9 & 7.807 & - & - \\
         & & 1266.9 & 7.893 & - & - \\
         & 34 & 1322.8 & 7.560 & 31.3 & 0.179 \\
         & & 1280.6 & 7.809 & - & - \\
         & & 1266.4 & 7.896 & - & - \\
         \hline
         \multirow{6}{*}{NH$_{3}$$^{P}$} & 15 & 1337.0 & 7.479 & 26.5 & 0.148 \\
         & 51 & 1337.3 & 7.478 & 25.3 & 0.142 \\
         & 60 & 1337.5 & 7.477 & 24.6 & 0.138 \\
         & 90 & 1337.8 & 7.475 & 23.4 & 0.131 \\
         & 99 & 1337.5 & 7.477 & 24.1 & 0.135 \\
         & 108 & 1336.3 & 7.483 & 27.7 & 0.155 \\
        \hline
        \multirow{6}{*}{H$_{2}$O:CO$_{2}$$^{M}$} & 15 & 1340.9 & 7.458 & 38.0 & 0.211 \\
         & 51 & 1341.4 & 7.455 & 35.3 & 0.196 \\
         & 101 & 1342.6 & 7.448 & 31.7 & 0.176 \\
         & 131 & 1341.9 & 7.452 & 30.1 & 0.167 \\
         & 149 & 1338.2 & 7.473 & 38.0 & 0.211 \\
         & 161 & 1334.6 & 7.493 & 36.4 & 0.204 \\
        \hline
        \multirow{6}{*}{H$_{2}$O:NH$_{3}$$^{P}$} & 15 & 1338.7 & 7.470 & 38.1 & 0.212 \\
         & 51 & 1340.2 & 7.462 & 35.4 & 0.198 \\
         & 99 & 1339.9 & 7.463 & 31.6 & 0.176 \\
         & 134 & 1339.0 & 7.468 & 30.9 & 0.173 \\
         & 149 & 1331.2 & 7.512 & 31.1 & 0.175 \\
         & 161 & 1330.8 & 7.514 & 27.7 & 0.157 \\
        \hline
        \multirow{3}{*}{CO:NH$_{3}$$^{M}$} & 15 & 1327.6 & 7.532 & 28.4 & 0.162 \\
         & 30 & 1330.3 & 7.517 & 25.6 & 0.144 \\
         & 37 & 1331.7 & 7.509 & 25.6 & 0.144 \\
        \hline
        
\end{tabular}
\end{center}
\label{cn_peaks_fwhm}
\end{table}

\clearpage

\section{Relative apparent band strengths of formamide in pure and mixed ices}
\label{appendix_b}

This appendix provides the relative apparent band strengths (\textit{$\eta$}) of formamide, calculated via Equation~\ref{eq:relbs}, where the value of \textit{A'} used in the calculations is the apparent band strength of the respective band in the pure amorphous formamide ice at 15 K (given in Table~\ref{band-strength}). Thus, the \textit{$\eta$} value of each band in the pure ice at 15 K is unity. The integration ranges used to calculate the integrated absorbances are stated for each mixture individually, as the same integration ranges were not used for all mixtures due to shifting peak positions and FWHMs. Different integration ranges were used to calculate the integrated absorbances of the amorphous and crystalline pure formamide peaks for the same reason.

These \textit{$\eta$} values can be used to calculate column densities or upper limits of formamide in a specific mixture and at a specific temperature by simply multiplying the corresponding relative apparent band strength by the appropriate apparent band strength in Table~\ref{band-strength}.

\begin{table*}[h]
\caption[]{Relative apparent band strengths of the three selected bands of pure formamide and formamide in various binary and tertiary mixtures at temperatures from 15-212 K.}
\fontsize{9}{10}\selectfont 
\begin{center}
\begin{tabular}{|c|c|c|c|c|}
\hline
         \multirow{3}{*}{Matrix} & \multirow{3}{*}{T (K)} & Rel. Band Strength & Rel. Band Strength & Rel. Band Strength \\
           &  & C=O stretch + NH$_{2}$ bend & CH bend & CN stretch \\
           &  & 1700.3 cm$^{-1}$ (5.881 $\mu$m) & 1388.1 cm$^{-1}$ (7.204 $\mu$m) & 1700.3 cm$^{-1}$ (7.529 $\mu$m) \\
        \hline 
         \multirow{9}{*}{Pure} &  & Int. Range: 2050-1532 cm$^{-1}$ & Int. Range: 1532-1364 cm$^{-1}$ & Int. Range: 1364-1258 cm$^{-1}$ \\\cline{3-5}
         & 15 & 1.00 & 1.00 & 1.00 \\
         & 49 & 0.99 & 1.00 & 0.97 \\
         & 99 & 0.96 & 0.99 & 0.93 \\
         & 149 & 0.95 & 0.97 & 0.91 \\\cline{3-5}
         &  & Int. Range: 2050-1570 cm$^{-1}$ & Int. Range: 1570-1364 cm$^{-1}$ & Int. Range: 1364-1258 cm$^{-1}$ \\\cline{3-5}
         & 170 & 0.70 & 1.01 & 1.015 \\
         & 191 & 0.70 & 1.01 & 1.014 \\
         & 212 & 0.60 & 0.82 & 0.857 \\
         \hline
        \multirow{7}{*}{H$_{2}$O} &  & Int. Range: 1961-1553 cm$^{-1}$ & Int. Range: 1532-1368 cm$^{-1}$ & Int. Range: 1368-1258 cm$^{-1}$ \\\cline{3-5}
         & 15 & 1.64 & 1.63 & 1.08 \\
         & 49 & 1.66 & 1.67 & 1.13 \\
         & 99 & 1.71 & 1.61 & 1.16 \\
         & 129 & 1.63 & 1.57 & 1.18 \\
         & 149 & 1.73 & 2.22 & 2.01 \\
         & 161 & 1.32 & 2.10 & 2.00 \\
         \hline
        \multirow{7}{*}{CO$_{2}$} &  & Int. Range: 1961-1532 cm$^{-1}$ & Int. Range: 1532-1364 cm$^{-1}$ & Int. Range: 1364-1240 cm$^{-1}$ \\\cline{3-5}
         & 15 & 1.33 & 0.85 & 1.94 \\
         & 34 & 1.34 & 0.84 & 1.88 \\
         & 40 & 1.31 & 0.78 & 1.85 \\
         & 49 & 1.31 & 0.79 & 1.88 \\
         & 60 & 1.33 & 0.83 & 1.92 \\
         & 90 & 1.36 & 0.93 & 1.91 \\
         \hline
        \multirow{4}{*}{CO} &  & Int. Range: 1900-1500 cm$^{-1}$ & Int. Range: 1532-1364 cm$^{-1}$ & Int. Range: 1364-1240 cm$^{-1}$ \\\cline{3-5}
         & 15 & 1.81 & 1.30 & 2.62 \\
         & 31 & 1.81 & 1.34 & 2.71 \\
         & 34 & 1.80 & 1.38 & 2.72 \\
         \hline
        \multirow{7}{*}{CH$_{3}$OH} &  & Int. Range: 1800-1550 cm$^{-1}$ & - & - \\\cline{3-5}
         & 15 & 1.42 & - & - \\
         & 50 & 1.41 & - & - \\
         & 100 & 1.39 & - & - \\
         & 121 & 1.75 & - & - \\
         & 139 & 1.70 & - & - \\
         & 154 & 0.79 & - & - \\
         \hline
        \multirow{7}{*}{NH$_{3}$} &  & - & Int. Range: 1484-1374 cm$^{-1}$ & Int. Range: 1374-1258 cm$^{-1}$ \\\cline{3-5}
         & 15 & - & 1.28 & 1.14 \\
         & 51 & - & 1.28 & 1.11 \\
         & 60 & - & 1.28 & 1.10 \\
         & 90 & - & 1.24 & 1.05 \\
         & 99 & - & 1.39 & 1.13 \\
         & 108 & - & 1.41 & 1.15 \\
         \hline
        \multirow{7}{*}{H$_{2}$O:CO$_{2}$} &  & Int. Range: 1961-1553 cm$^{-1}$ & Int. Range: 1437-1368 cm$^{-1}$ & Int. Range: 1368-1279 cm$^{-1}$ \\\cline{3-5}
         & 15 & 2.20 & 1.30 & 0.75 \\
         & 51 & 2.18 & 1.23 & 0.67 \\
         & 101 & 1.89 & 1.10 & 0.51 \\
         & 131 & 1.85 & 0.99 & 0.50 \\
         & 149 & 1.88 & 1.55 & 1.24 \\
         & 161 & 1.57 & 1.42 & 1.33 \\
         \hline
        \multirow{7}{*}{H$_{2}$O:NH$_{3}$} &  & - & Int. Range: 1437-1368 cm$^{-1}$ & Int. Range: 1368-1279 cm$^{-1}$ \\\cline{3-5}
         & 15 & - & 1.35 & 1.19 \\
         & 51 & - & 1.36 & 1.21 \\
         & 99 & - & 1.32 & 1.27 \\
         & 134 & - & 1.19 & 1.27 \\
         & 149 & - & 1.11 & 1.24 \\
         & 161 & - & 1.01 & 1.22 \\
         \hline
        \multirow{4}{*}{CO:CH$_{3}$OH} &  & Int. Range: 1810-1524 cm$^{-1}$ & - & - \\\cline{3-5}
         & 15 & 1.57 & - & - \\
         & 30 & 1.64 & - & - \\
         & 35 & 1.68 & - & - \\
         \hline
        \multirow{4}{*}{CO:NH$_{3}$} &  & - & Int. Range: 1460-1370 cm$^{-1}$ & Int. Range: 1370-1256 cm$^{-1}$ \\\cline{3-5}
         & 15 & - & 1.41 & 2.23 \\
         & 30 & - & 1.49 & 2.14 \\
         & 37 & - & 1.54 & 1.85 \\
         \hline
        
\end{tabular}
\end{center}
\label{rel_band_strengths}
\end{table*}

\clearpage

\section{QMS calibration of an independent tri-dosing leak valve system and mixing ratio determination}
\label{supp_info_qms}

\subsection{Calibration procedure and mixing ratio determination}

The new tri-dosing system mentioned in Section~\ref{section:methodology} allows for simultaneous but independent deposition of gases and vapors via three leak valves, each connected to a separate gas line. Compared to our previous method in which gases and vapors were premixed in the desired ice ratio in a gas bulb and then dosed into the chamber through a single valve, the new method allows for codepositing multiple gases and vapors without experimental errors in the ratio caused by mixing gases with different volatilities in a single bulb or dosing gases that may have different flow, pumping, and substrate deposition rates through the same valve. Subsequently, it greatly improves the ability to create mixtures with precisely determined ratios of molecules with low volatilities like formamide, which is challenging in traditional premixing procedures. The benefits of independent multidosing systems were also described for similar systems with two leak valves in \citet{gerakines1995infrared} and \citet{yarnall2022new}.

There are several ways to calibrate such a system to ensure a certain ratio of ice components. One such method is calibrating the deposition rate on the substrate to a specific leak valve position with a specific pressure of the gas or vapor of choice in its manifold line. However, because formamide has a very low vapor pressure compared to liquids like H$_{2}$O and CH$_{3}$OH and tends to stick to and condense in various parts of the line, reproducing a specific line pressure throughout multiple experiments using this method is difficult. Therefore, to conduct a systematic and thorough IR characterization of formamide in a wide variety of ices with precisely constrained mixing ratios, a different method is necessary.

For this purpose, we calibrate molecules' ice deposition rates with the intensity of their mass signals during the deposition with a QMS. In this calibration procedure, a pure molecule is dosed at a constant rate into the chamber, with the substrate cooled to the desired deposition temperature and the IR spectrometer continuously collecting IR spectra, while the QMS continuously collects mass peak intensity values of selected mass-to-charge ratios (m/z) in the selected ion monitoring (SIM) mode. The IR spectrometer is used to measure the ice column density rather than the laser interference because the formamide deposition pressure does not remain stable over the long period of time necessary to generate multiple interference fringes ($>$18 hours), which is necessary to reliably extract a deposition rate. Conversely, a deposition rate can be extracted from integrated absorbance growth rates (obtained via a least-squares fit to the integrated absorbance over time) in $\sim$30 mins, during which time the formamide deposition rate remains stable (as indicated by the linearity of the integrated absorbance increase over time). The integrated absorbance growth rate for that molecule can then be correlated to a specific mass peak's signal intensity (typically the molecule's base peak) in the QMS (obtained via averaging the mass peak's signal intensity values collected during the deposition and simultaneous IR data collection). The integrated absorbance growth rate can then be converted to the ice column density growth rate, \textit{dN/dt}, via the following equation if the band strength of the pure molecule, \textit{A}, is known:

\begin{equation}
    \frac{dN}{dt} = \frac{2.303}{A} \times \frac{d\int abs(\nu) \ d\nu}{dt}
\label{eq:dndt}
.\end{equation}

\begin{figure*}
\centering
\includegraphics[width=\hsize]{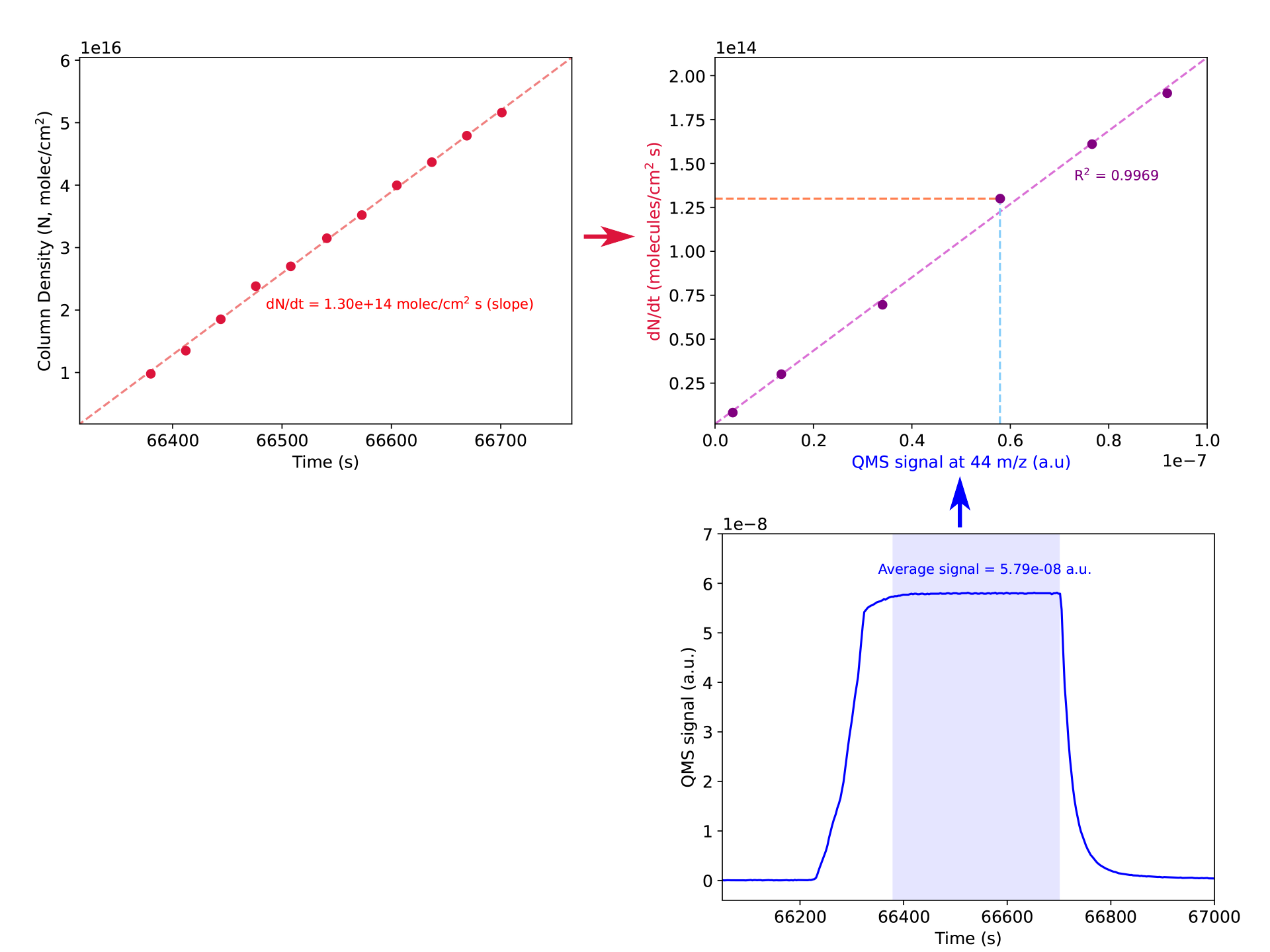}
\caption{Example of how a calibration curve is obtained, shown here for CO. The rate of ice growth is found via monitoring the molecule's integrated absorbance of a selected peak, which can be converted to column densities via known pure band strengths, and relating them to the time of the spectrum collected (top left). Then the average value of a given mass signal (28 m/z for CO), measured by the QMS during the same time interval as the collection of the IR spectra, is calculated (bottom right). The calibration curve is obtained by correlating the two values to each other for a range of ice deposition rates (top right).}
\label{fig:cal_curve_method}
\end{figure*}

Table~\ref{ratios} provides the peak used for the calibration of each pure molecule and its corresponding band strength and reference.

Via this method, a calibration curve relating a mass peak's signal intensity in the QMS to its column density growth rate can be determined, with the slope of this curve referred to here as a molecule's \textit{sensitivity} (see Figure~\ref{fig:cal_curve_method} for an example of such a calibration). When starting a deposition, the leak valve can then be opened accordingly so that the mass signal of the molecule in the QMS corresponds to the desired column density growth rate. In this work, such calibration curves were completed for all molecules used in these spectra with a Spectra Microvision Plus QMS. The relationship between column density growth rate and QMS signal intensity is linear for all molecules within the deposition pressure ranges used (R$^{2}$ values of the linear fits ranged from 0.9699-0.9999 with an average of 0.9936).

\begin{table*}[h]
\caption{Peaks and their band strengths used to obtain calibration curves of the pure molecules.}
\begin{center}
\begin{tabular}{c  c  c}
\hline
        Molecule & Calibration peak & Band strength used \\
        & (cm$^{-1}$ | $\mu$m) & (10$^{-17}$ cm molec$^{-1}$) \\
        \hline
        NH$_{2}$CHO & 1701 | 5.88 & 6.4 (this work) \\
        H$_{2}$O$^{*}$ & \begin{tabular}{c}3280 | 3.05 \\ 1660 | 6.02 \end{tabular} & \begin{tabular}{c}20. \citep{bouilloud2015bibliographic} \\ 1.1 \citep{bouilloud2015bibliographic} \end{tabular} \\
        CO$_{2}$$^{*}$ & \begin{tabular}{c}2343 | 4.27 \\ 665 | 15.3 \end{tabular} & \begin{tabular}{c}13. \citep{bouilloud2015bibliographic} \\ 1.8 \citep{bouilloud2015bibliographic} \end{tabular} \\
        CO & 2139 | 4.67 & 0.87 \citep{gonzalez2022density} \\
        CH$_{3}$OH$^{*}$ & \begin{tabular}{c}1027 | 9.74 \\ 1460 | 6.85 \end{tabular} & \begin{tabular}{c}1.61 \citep{luna2018densities} \\ 1.1 \citep{luna2018densities} \end{tabular} \\
        NH$_{3}$ & 1071 | 9.34 & 1.95 \citep{hudson2022ammonia} \\
    \hline
     \label{a-val-cal}
\end{tabular}

\begin{tablenotes}
    \item[\emph{}]{$^*$ If two calibration peaks are listed, then both peaks were utilized to create the calibration curve, with the peak with a lower band strength used to calculate the ice's column density at the higher deposition rates to avoid saturation effects in the calibration curve.}

\end{tablenotes}

\end{center}
\end{table*}

After the experiment, the mass signal data during the deposition can be converted via the equation from the calibration curve to a column density growth rate, which is then integrated over time to give the absolute column density of each species at the end of the deposition. However, in the case that some of the species in a given mixture share their strongest mass peaks and have no alternative strong peaks without overlap with the other mixture components (which is the case for several mixtures in this work), the individual column density growth rates must be extracted from the mass spectra by utilizing ratios of a given molecule's base peak to another mass peak that is not shared with any other molecules in a given mixture. For example, the mass spectrum of formamide contains a peak at 28, the base peak of CO. Thus, during the deposition of the formamide:CO mixture, the 28 m/z signal contains contribution from both formamide and CO. The contribution of formamide to the signal at 28 m/z was calculated by dividing the signal at 45 m/z (which, in this mixture, only formamide contributed to) by the ratio of the 45 and 28 m/z peaks during pure formamide deposition. This calculated contribution was then subtracted from the 28 m/z signal to yield the CO 28 m/z signal.

In order to estimate the error of the calculated column density of each component and, subsequently, the mixing ratios in each ice, multiple sources of error have been considered. These are discussed in the following subsections.

\subsection{Ion interference effect}

Ion interactions within the instrument, such as ion-molecule interactions or ions interacting with the QMS filament or rods, during the dosing of multiple species into the chamber can effect a molecule's sensitivity. Such interactions between two different species can cause their sensitivities to deviate from the values determined in the calibration of each species in pure form. This phenomenon is often referred to as the \textit{ion interference effect}, and it complicates using a mass spectrometer to quantify gases or vapors in a mixture \citep{basford1993recommended,yongjun2022study}.

\begin{figure*}
\centering
\includegraphics[scale=0.7]{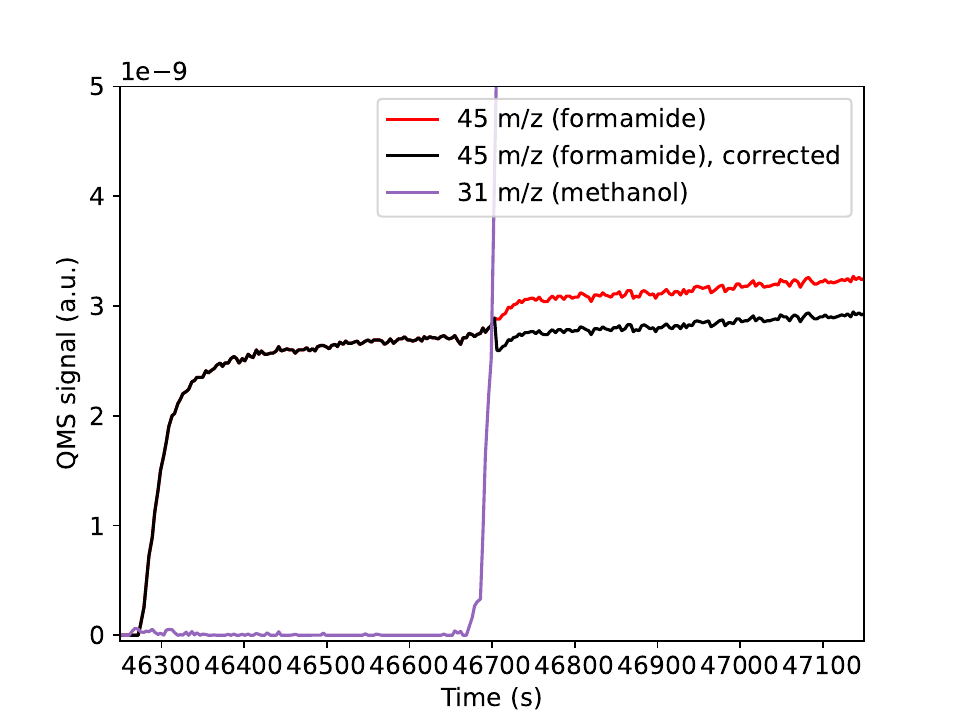}
\caption{Correction of the ion interference effect observed in the formamide 45 m/z signal during CH$_{3}$OH codeposition.}
\label{fig:ief_correction}
\end{figure*}

The magnitude of this effect is highly dependent on the species as well as the instrument. It increases with total pressure and decreases for a given species as its proportion in a mixture increases \citep{nemanivc2018argon,sun2020study}. Thus, the sensitivities that are most affected by this effect are those of species that are present in the lowest proportions in a mixture. Given that our formamide dosing pressure was in the range of a couple 10$^{-9}$ mbar and that the intended ratio of formamide to matrix components was $\sim$5:100 in the case of binary mixtures and 5:100:25 in the case of tertiary mixtures, we treated the interference effect of formamide on the matrix components as negligible and accounted for ion interference only in the formamide signal. While the formamide absolute column densities are necessary to calculate its relative band strengths (see Section \ref{section:methodology}), the absolute column densities of the matrix components are not needed to find any values other than the mixing ratios.

In order to quantify the ion interference effect on formamide in each mixture, at the start of each deposition, formamide was first dosed alone, and its mass signal was given $\sim$5 min to stabilize before the other matrix components were introduced into the chamber. Although this meant that each experiment started with a very brief deposition of pure formamide, the deposition rate of formamide was so slow in all of the experiments (on the order of tens of monolayers per hour) that this brief pure deposition was usually not even noticeable above the noise level in the IR spectra. Then, the ratio between formamide's signal before and after the matrix molecules were added to the chamber was used as a correction factor to remove the ion interference effect from formamide's signal. An example of this correction is shown in Figure~\ref{fig:ief_correction} for the formamide:CH$_{3}$OH mixture, which had the highest correction factor of all the mixtures (1.11).

The ion interference effect on formamide was noticeable in all of the mixtures where the major matrix component was polar, while it was not detected above the noise level in the mixtures in which the major matrix component was apolar. In order to provide a conservative estimate of the error caused by the ion interference effect on the calculated column density of formamide, the percent difference of the formamide column density before and after the ion interference effect correction was obtained for all mixtures in which the effect was detected above the noise level. The average percent difference was $\sim$5\%, with the highest percent difference being that of the formamide:CH$_{3}$OH mixture ($\sim$10\%). To avoid underestimating the error in any of the mixtures, we use this maximum error, $\sim$10\%, as the uncertainty in the column density caused by the ion interference effect.

The sensitivity of a QMS can also drift over time. However, such drift is typically only significant over timescales spanning several months to a couple years, and it is more significant for absolute sensitivities than for relative sensitivities \citep{ellefson1987calibration,lieszkovszky1990metrological}. The contribution of this drift to the method error was assumed to be negligible here given that all the formamide mixture spectra were collected within a span of two months, and that the calibration curves were usually either determined within a few days of their use to create an ice mixture or were frequently updated with new values that were consistent with the fits to the previous values.

\subsection{Error calculation}

In this method of determining ice column densities, multiple sources of error must be considered. First, there is the error in the method used to determine the ice column density growth rate (\textit{dN/dt}). This error can be estimated by propagating uncertainties of the variables in Equation~\ref{eq:dndt}. For all ices, the integrated absorbance growth rate uncertainty is estimated to be 10\%, as mentioned in Section~\ref{section:methodology}. For formamide, the uncertainty in the band strengths reported in this work is estimated to be 15\% (also see Section~\ref{section:methodology}). For the matrix components, literature band strength values were used (see Table~\ref{a-val-cal}). However, in the literature, variations between the band strengths reported in different publications can be large (e.g., \citealt{bouilloud2015bibliographic}). For this reason, we estimate a 25\% uncertainty for the literature band strengths used.

Then, we determine the uncertainty resulting from converting the integrated QMS measurement to a column density experimentally by finding the difference between the column density calculated from the QMS signal and the column density calculated from the integrated IR absorbance at the end of a pure molecule's IR measurement. Comparing these differences in formamide, H$_{2}$O, and CO deposition experiments resulted in an average error $\sim$2.5\%. To be conservative, we estimate the error from converting the QMS measurement to a column density to be 5\%.

Propagating all of these uncertainties, along with the 10\% uncertainty from the ion interference effect for the formamide measurements, results in an uncertainty of $\sim$21\% for the formamide column densities and $\sim$27\% for the matrix column densities in the ice mixtures.

\end{appendix}

\end{document}